\documentclass[11pt]{article}

\usepackage{beton}
\usepackage{euler}
\usepackage{sfbold}
\usepackage{fullpage}
\usepackage{array}
\usepackage[dvips]{graphicx}
\usepackage{amsmath}
\usepackage{amsthmlike}
\usepackage{thrmlist}
\usepackage{multicol}
\usepackage{version}
\usepackage{psfrag}
\includeversion{longonly}
\excludeversion{shortonly}

\newcommand{\squash}[1]{\makebox[0pt][c]{#1}}

\newcommand{\set}[1]{\{\,#1\,\}}
\DeclareMathOperator{\CH}{conv}
\DeclareMathOperator{\Wedge}{wedge}
\DeclareMathOperator{\kernel}{kernel}

\DeclareMathOperator{\interior}{int}

\newcommand{\hexlabel}{%
  \psfrag{a}{$a$}
  \psfrag{b}{$b$}
  \psfrag{c}{$c$}
  \psfrag{d}{$d$}
  \psfrag{e}{$e$}
  \psfrag{f}{$f$}
  \psfrag{s}{$s$}
  \psfrag{i}{$i$}
  \psfrag{r}{$r$}
  \psfrag{ap}{$a'$}
  \psfrag{ep}{$e'$}
}

\newlength{\figwidth}
\def\Caption#1#2{\centerline{\begin{minipage}[b]{0.8\textwidth}%
\normalfont\fontsize{10}{12}\selectfont
\caption{#1}\label{#2}\end{minipage}}}

\newcommand{\email}[1]{\texttt{#1}}
\newcommand{\support}[1]{\quad #1}

\title{Small Strictly Convex Quadrilateral Meshes of Point Sets}

\author{
  David Bremner\footnotemark[1]{}
\and Ferran Hurtado\footnotemark[2]{}
\and Suneeta Ramaswami\footnotemark[3]{}
\and Vera Sacrist\'an\footnotemark[2]{}}

\begin{shortonly}
  \institute{Faculty of Computer Science, University of New
    Brunswick\thanks{Email: \email{bremner@unb.ca}. Partially
      supported by NSERC Grant RGPIN 228095.}  \and Dep. Matem\`atica
    Aplicada II, Universitat Polit\`ecnica de Catalunya\thanks{Email:
      \email{\{hurtado,vera\}@ma2.upc.es}. Partially supported by CUR
      Gen.\ Cat.\ 1999SGR00356 and Proyecto DGES-MEC PB98-0933.}  \and
    Dep. Computer Science, Rutgers University, Camden\thanks{ Email:
      \email{rsuneeta@camden.rutgers.edu}.  Partially supported by a
      Rutgers University Research Council Grant URF-G-00-01.}}
\end{shortonly}

\date{\today}

\begin{document}
{
\renewcommand{\thefootnote}{\fnsymbol{footnote}}
\footnotetext[1]{
 Faculty of Computer Science, University of New Brunswick,
P.O. Box 4400, Fredericton, NB, E3B 5A3 Canada.
\email{bremner@unb.ca} 
\support{Partially supported by an NSERC Individual Research Grant}}

\footnotetext[2]{
 Dep.\ Matem\`atica Aplicada II, Universitat Polit\`ecnica de
Catalunya, Pau Gargallo 5, 08028 Barcelona, Spain.              
\email{ \{ hurtado, vera \}@ma2.upc.es}
\support{Partially supported by CUR Gen.\ Cat.\ 1999SGR00356 and
Proyecto DGES-MEC  PB98-0933.}}

\footnotetext[3]{
   Dep.\ Computer Science, 322 Business and Science Building, 
   Rutgers University, Camden, NJ 08102,
   USA. \email{rsuneeta@camden.rutgers.edu}} 
}
\maketitle

\thispagestyle{empty}

\begin{abstract}
In this paper, we give upper and lower bounds on the number of Steiner
points required to construct a strictly convex quadrilateral mesh for
a planar point set. In particular, we show that
$3{\lfloor\frac{n}{2}\rfloor}$ internal Steiner points are always
sufficient for a convex  
quadrilateral mesh of $n$ points in the plane. Furthermore, for any
given 
$n\geq 4$, there are point sets for which
$\lceil\frac{n-3}{2}\rceil-1$ Steiner points are necessary for a
convex 
quadrilateral mesh.

\end{abstract}

\section{Introduction}

Discrete approximations of a surface or volume are necessary in
numerous applications. Some examples are models of human organs in
medical imaging, terrain models in GIS, or models of parts in a
CAD/CAM system.  These applications typically assume that the
geometric domain under consideration is divided into small, simple
pieces called {\em finite elements}. The collection of finite elements
is referred to as a {\em mesh}.  For several applications, {\em
quadrilateral/hexahedral} mesh elements are preferred over
triangles/tetrahedra owing to their numerous benefits, both geometric and
numerical; for example, quadrilateral meshes give lower approximation
errors in finite element methods for elasticity
analysis~\cite{al88,bpmcs} or metal forming
processes~\cite{jsk91}. However, much less is known about
quadrilateralizations and hexahedralizations and in general,
high-quality quadrilateral/hexahedral \begin{longonly}(quad/hex)\end{longonly} meshes are harder to 
generate than good triangular/tetrahedral \begin{longonly}(tri/tet)\end{longonly} ones. 
  Indeed, there
are several important open questions, both combinatorial as well as
algorithmic, about quad/hex meshes for sets of objects such as
polygons, points, etc., even in two dimensions.
Whereas triangulations of polygons and two-dimensional (2D) point sets
and tetrahedralizations of three-dimensional (3D) point sets and
convex polyhedra always exist (not so for non-convex
polyhedra~\cite{schon}), quadrilateralizations of 2D point sets do
not. Hence it becomes necessary to add extra points, called {\em
  Steiner points},
to the geometric domain.  This
raises the issue of bounding the number of Steiner points, and hence
the mesh complexity, while also providing guarantees on the quality of
element shape.  
Such problems are especially relevant for applications in scattered
data interpolation~\cite{cl00,lai00,LS97}, which require quadrilateral
meshes that modify the original data as little as possible, i.e., add
few Steiner points.

A theoretical treatment of  quadrilateral/hexahedral meshes has only
recently begun~\cite{be97,bt97,epp96,mitchell2,mitchell1,mw97,rrt97}.
Some work on 
  quadrangulations\footnote{In this paper, we use the term 
{\em quadrangulation} 
interchangeably with quadrilateralization. Both terms are common 
in the meshing literature.}
 of restricted classes of polygons has been done
in the computational geometry community~\cite{eow84,kkk83,lu85,sa82}.
However, there are numerous unresolved questions. For example, even
the fundamental question of deciding if a 2D set of points admits a
convex quadrangulation without the addition of Steiner points, is
unsolved.  A survey of results on quadrangulations of planar sets
appears in~\cite{toussaint95}.

Any planar point set can be quadrangulated with at most one Steiner
point, which is required only if the number of points on the convex
hull is odd~\cite{bt97}. For planar simple $n$-gons, $\lfloor
n/4\rfloor$ internal Steiner points suffice to quadrangulate the
polygon~\cite{rrt97}.  In both cases, the quadrilaterals of the
resulting mesh will be, in general, non-convex.  However, for many
applications, an important requirement is that the quadrangulation be
{\em strictly convex}, {i.e.}, every quadrilateral of the mesh
must have interior angles strictly less than $180^{\circ}$. A natural
problem then is to construct strictly convex quadrilateral meshes for
planar geometric domains, such as polygons or point sets, with a
bounded number of Steiner points.  Some results on convex
quadrangulations of planar simple polygons are known. For example, it
was shown in~\cite{elosu} that any simple $n$-gon can be decomposed
into at most $5(n-2)/3$ strictly convex quadrilaterals and that $n-2$
are sometimes necessary.  Furthermore, circle-packing
techniques~\cite{be97,bmr95,be92a} have been used to generate, for a
simple polygon, quadrilateral meshes in which no quadrilateral has
angle greater than $120^\circ$. For planar point sets, experimental
results on the use of some heuristics to construct quadrangulations
with many convex quadrangles appear in~\cite{brtt}. In~\cite{fmr}, it
is shown that a%
\begin{longonly}
related optimization problem, namely%
\end{longonly}
finding a minimum weight
convex quadrangulation (i.e.\ where the sum of the edge lengths is
minimized) can be found in polynomial time for point sets
constrained to lie on a fixed number of convex layers.

In this paper, we study the problem of constructing a strictly convex
quadrilateral mesh for a planar point set using a bounded number of
Steiner points.  We use ``convex-quadrangulate'' to mean ``obtain a 
strictly convex quadrangulation for''. If the number of extreme points 
of the set is even,
it is always possible to convex-quadrangulate the set using Steiner
points which are all internal to the convex hull. If the number of
points on the convex hull is odd, the same is true, assuming that in
the quadrangulation we are allowed to have exactly one triangle.  We
provide upper and lower bounds on the number of Steiner points
required for a strictly convex quadrangulation of a planar point set.  In
particular, in Section~\ref{sec:lower}, we prove that for any $n\geq
4$, $\lceil\frac{n-3}{2}\rceil-1$ Steiner points may sometimes be
necessary to convex-quadrangulate a set of $n$ points.  In
Section~\ref{sec:upper}, we prove that $3{\lfloor\frac{n}{2}\rfloor}$
internal Steiner points are always sufficient to convex-quadrangulate
any set of $n$ points.

\section{Lower bound\label{sec:lower}}

In this section we describe a particular configuration of $m+3\geq 4$
points which requires at least $\lceil{\frac{m}{2}}\rceil-1$
Steiner points to be convex-quadrangulated. We also show a
convex-quadrangulation of the set that uses close to that few Steiner
points.

\paragraph{Description of the configuration of points:} 
The configuration of $m+3$ points consists of $m+1$ points placed
along a line $\ell$, with one point above the line and another point
below the line, such that the convex hull of the set has 4 vertices,
namely the extreme points on the line and the top and bottom points
(see Figure \ref{fig:qline1}). We refer to the vertices on $\ell$ as
{\em line vertices}. We will refer to the entire configuration as $S$.
\begin{figure}[tbhp]                                                        
\centerline{\includegraphics[width=.6\textwidth]{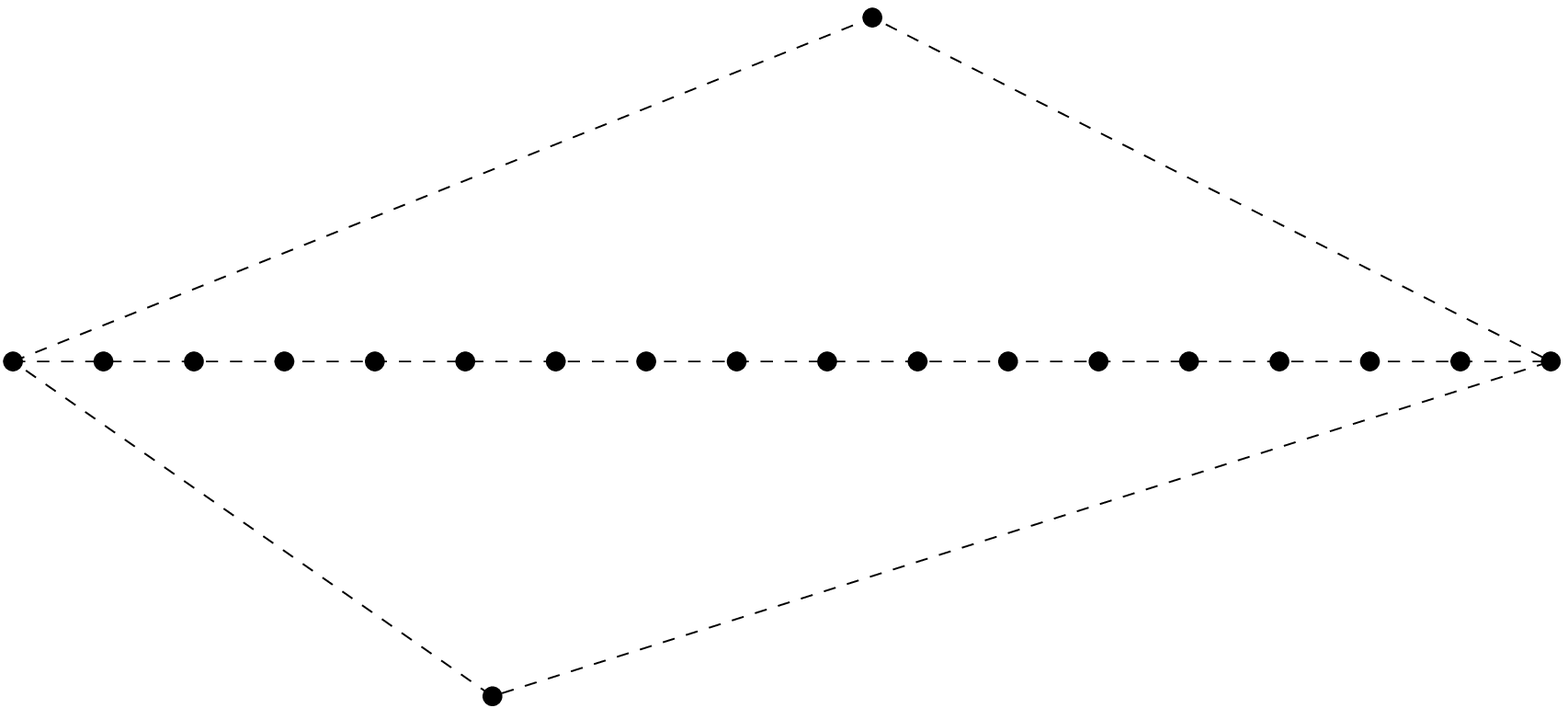}}
\Caption{The point set $S$ has $m+1$ points along the line, plus the
top and the bottom points. Its convex hull is a quadrangle.}
{fig:qline1}
\end{figure}

Consider any strictly convex quadrangulation $\cal C$ of the
set. Since all the quadrangles in $\cal C$ are strictly convex, each
point on $\ell$ must belong to at least one edge of the
quadrangulation lying strictly above the line, and at least one edge
lying strictly below the line. Quadrangulation edges incident on an
input point and lying above (below) $\ell$ will be called {\em upward}
({\em downward}) edges.

\begin{figure}[tbhp]                                                        
\psfrag{a1}{$a_1$}
\psfrag{a2}{$a_2$}
\psfrag{am}{$a_m$}
\psfrag{u1}{$u_1$}
\psfrag{u2}{$u_2$}
\psfrag{d1}{$d_1$}
\psfrag{d2}{$d_2$}
\centerline{\includegraphics[width=.7\columnwidth]{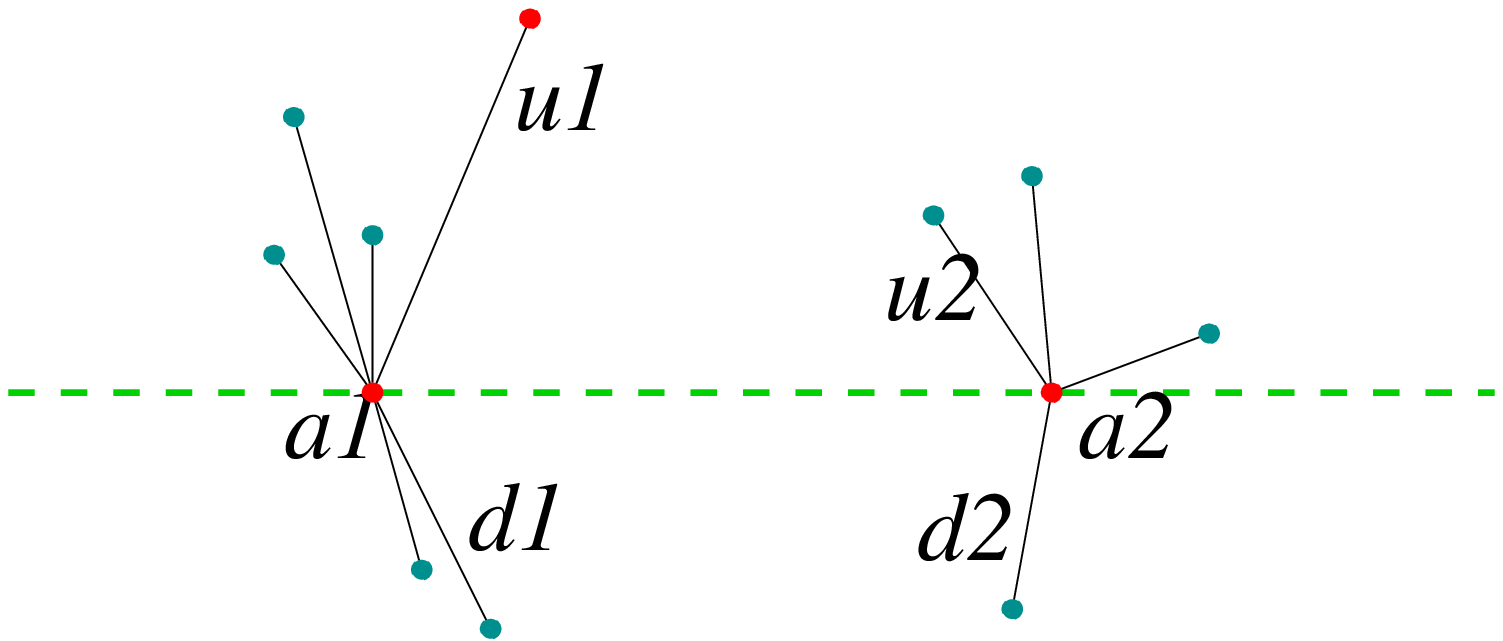}}
\caption{Relevant upward and downward edges.}
\label{fig:qline2}
\end{figure}
\begin{figure}

\centerline{\includegraphics[width=.8\columnwidth]{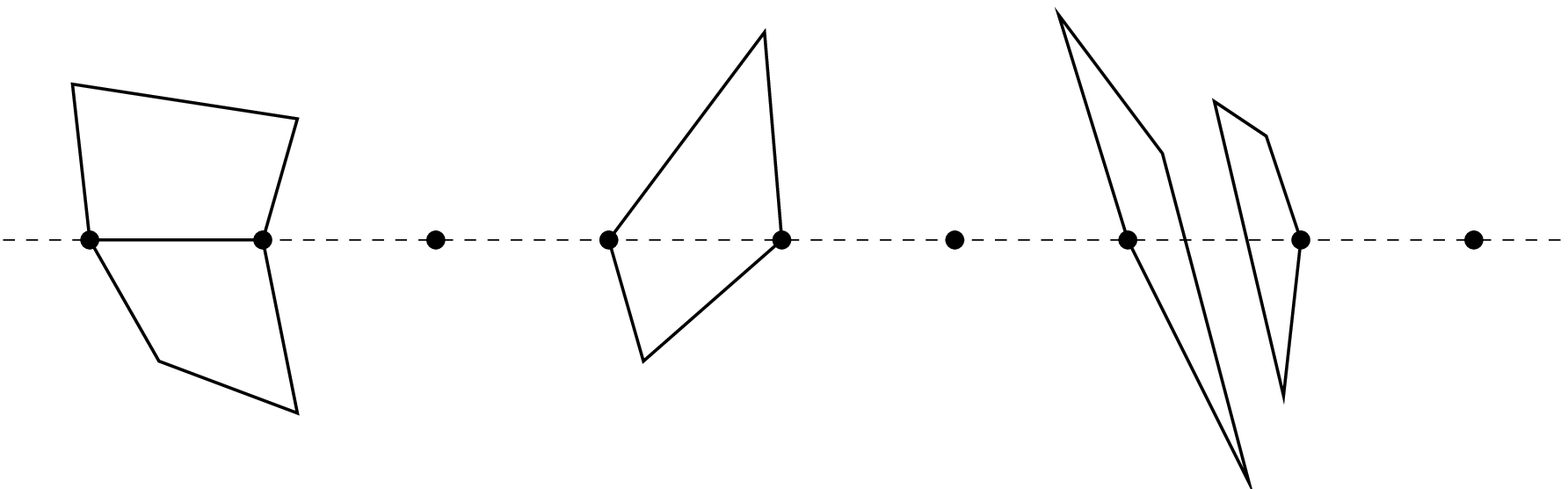}}
\caption{Squares, diamonds and half-diamonds.}
\label{fig:qline3}
\end{figure}

Consider two consecutive points $a_1$ and $a_2$ on $\ell$ with $a_1$
to the left of $a_2$.  Let $u_1$ be the clockwise last upward edge
incident on $a_1$, and let $u_2$ be the counterclockwise last upward
edge incident on $a_2$.  Symmetrically, let $d_1$ be the
counterclockwise last downward edge incident on $a_1$ and let $d_2$ be
the clockwise last downward edge incident on $a_2$ (see Figure
\ref{fig:qline2}).
If $(a_1, a_2)$ is an edge of $\cal C$, then it must form one
quadrangle of $\cal C$ together with $u_1$ and $u_2$, and another one
with $d_1$ and $d_2$. We call these two faces \emph{squares}.  If
$(a_1, a_2)$ is not an edge of $\cal C$, $u_1$ and $d_1$ must belong
to the same quadrangle, and so must also $u_2$ and $d_2$. If these two
quadrangles are the same, we call it a \emph{diamond}. If they are
different, we call them a \emph{pair of half-diamonds}. These three
cases are illustrated in Figure~\ref{fig:qline3}.

\begin{theorem}
The point set $S$ requires at least $\lceil{\frac{m}{2}}\rceil-1$
Steiner points to be convex-quadrangulated. 
\label{theorem:line}
\end{theorem}

\begin{proof}
  \begin{shortonly}
    \emph{(Sketch)}
  \end{shortonly}
Consider the graph $G=(V,E)$ formed by taking the union of all the
squares, diamonds and half-diamonds, together with the convex hull
edges. This graph, which is a subgraph of $\cal C$, is planar and its
faces consist of the squares, the diamonds, the half-diamonds, and
possibly some other faces that we will call ``extra faces''.  Its
edges are all square, diamond, half-diamond, or convex hull edges.
Let $q$ be the number of squares, $d$ the number of diamonds and $h$
the number of half-diamonds. We have
\begin{equation}
  \label{eq:m}
m={\frac{q}{2}}+d+{\frac{h}{2}}\,.  
\end{equation}
  Let 
$v$, $e$, $f$ denote the number of vertices, edges and faces of $G$.
Let $s$ be the number of vertices that did not belong to the original
set, i.e., the number of Steiner points in $\cal C$. Let $x$ be
the number of extra faces. We have $v\le m+3+s$ (because not every
Steiner point need be a vertex of $G$), and $f=q+d+h+x$.  Since $G$ is
planar, we can apply Euler's formula and \eqref{eq:m} as follows:
\begin{longonly}
\begin{align*}
v+f&=e+2\\
(m+3+s)+(q+d+h+x)&\geq e+2\\
s&\geq e-{\frac{3}{2}}q-2d-{\frac{3}{2}}h-x-1
\end{align*}
\end{longonly}
\begin{shortonly}
  \begin{equation*}
    s\geq e-{\frac{3}{2}}q-2d-{\frac{3}{2}}h-x-1
  \end{equation*}
\end{shortonly}
Now, if we can prove that 
\begin{equation}\label{eq:charging}
e\ge{\frac{7}{4}}q+{\frac{5}{2}}d+{\frac{7}{4}}h+x,
\end{equation}
we will obtain that 
\begin{align*}
s & \geq e-{\frac{3}{2}}q-2d-{\frac{3}{2}}h-x-1\\[0.3ex]
    &\geq({\frac{7}{4}}-{\frac{3}{2}})q+({\frac{5}{2}}-2)d+({\frac{7}{4}}-{\frac{3}{2}})h-1\\[0.3ex]  
    &={\frac{q}{4}}+{\frac{d}{2}}+{\frac{h}{4}}-1=
    {\frac{m}{2}}-1\\[0.3ex]
&\geq\left\lceil{\frac{m}{2}}\right\rceil-1 \qquad   \text{(because $s$
    must be an integer)}. 
\end{align*}

\begin{shortonly}
  \begin{equation*}
s\geq {q\over 4}+{d\over 2}+{h\over 4}-1={m\over 2}-1\geq\left\lceil{m\over
2}\right\rceil-1\,.
  \end{equation*}
\end{shortonly}
The general scheme to establish (\ref{eq:charging}) will be to
partition the edges of (quadrangles in) $G$ into three sets, and then
charge each edge to the faces bounded by the edge. The classification
of edges and the charging scheme are as follows:
\begin{itemize}
\item
Line edges: edges with both endpoints on the line $\ell$.  Each such
edge is shared by a pair of squares. Each square gets charged
${1/2}$.
\item  
Steiner edges: edges with neither endpoint on the line $\ell$.  Each
such edge charges ${1/2}$ to each of the faces that it bounds.
\item
Vertical edges: edges with exactly one endpoint on the line $\ell$.
\begin{itemize}
\item
If a vertical edge is shared by two diamonds, each diamond gets
charged~${1/2}$.

\item
If it is shared by a diamond and an extra face, the diamond gets
charged ${3/4}$ and the extra face gets charged ${1/4}$.

\item
If it belongs to a square or a half-diamond, the square or
half-diamond gets charged ${3/8}$, and the other face gets
charged ${5/8}$.
\begin{longonly}
  Notice that in this last case the total charge
  is less than 1 when the edge is shared by squares and/or
  half-diamonds.
\end{longonly}
\end{itemize}
\end{itemize}
\begin{longonly}

As a result, each face of $G$ gets charged in the following way:
\begin{itemize}
\item
Each extra face is charged at least $1$, since it has at least four
edges (recall that $G$ is a subgraph of the quadrangulation $\cal C$)
and is charged at least ${1/4}$ from each edge. 

\item
Each square is charged ${7/4}$: $1/2$ from its line edge, $1/2$ from
its Steiner edge, and $3/8$ from each of its two vertical edges.

\item
Each half-diamond is charged ${7/4}$: $1/2$ from each of its two
Steiner edges and $3/8$ from each of its two vertical edges.

\item 
Each diamond is charged at least $5/2$. Notice that if a
diamond $\alpha$ shares one upward vertical edge $(a_1,u_1)$ with
another diamond or half-diamond $\beta$, then it must share the
coincident downward edge $(a_1,d_1)$ with an extra face, since no
square, diamond or half-diamond could share it. This is because {\em
(i)} $\alpha$ obviously cannot share $(a_1,d_1)$ with $\beta$ because
of strict convexity, and {\em (ii)} $\alpha$ cannot share $(a_1,d_1)$
with any other diamond, square, or half-diamond face because such a
face would intersect $\beta$. 
For the same reasons, if a diamond shares a downward edge $(a_1,d_1)$
with another diamond or half-diamond, the coincident upward edge
$(a_1,u_1)$ must be shared with an extra face.  So, if the diamond is
adjacent to another diamond, it is charged at least
${\frac{1}{2}}+{\frac{3}{4}}={\frac{5}{4}}$ from the two edges incident on
that 
line vertex.
If it is adjacent to a pair of squares, it is charged
${\frac{5}{8}}+{\frac{5}{8}}={\frac{5}{4}}$. Any other combination
would charge more. In 
total, the diamond gets charged at least $5/2$.
\end{itemize}
This proves that
\begin{equation*}
e\ge\sum \mathrm{charges} = x+{\frac{7}{4}}q+{\frac{7}{4}}h+{\frac{5}{2}}d.    
\end{equation*}
\end{longonly}
\end{proof}
\begin{figure}[tbhp]                                  
  \psfrag{v1}{$v_1$}
  \psfrag{p1}{$p_1$}
  \psfrag{v2}{$v_2$}
  \psfrag{p2}{$p_2$}
  \psfrag{vm}{$v_{m+1}$}
  \psfrag{t}{$t$}
  \psfrag{b}{$b$}
  \psfrag{s}{$s$}
\centerline{\includegraphics[width=0.8\textwidth]{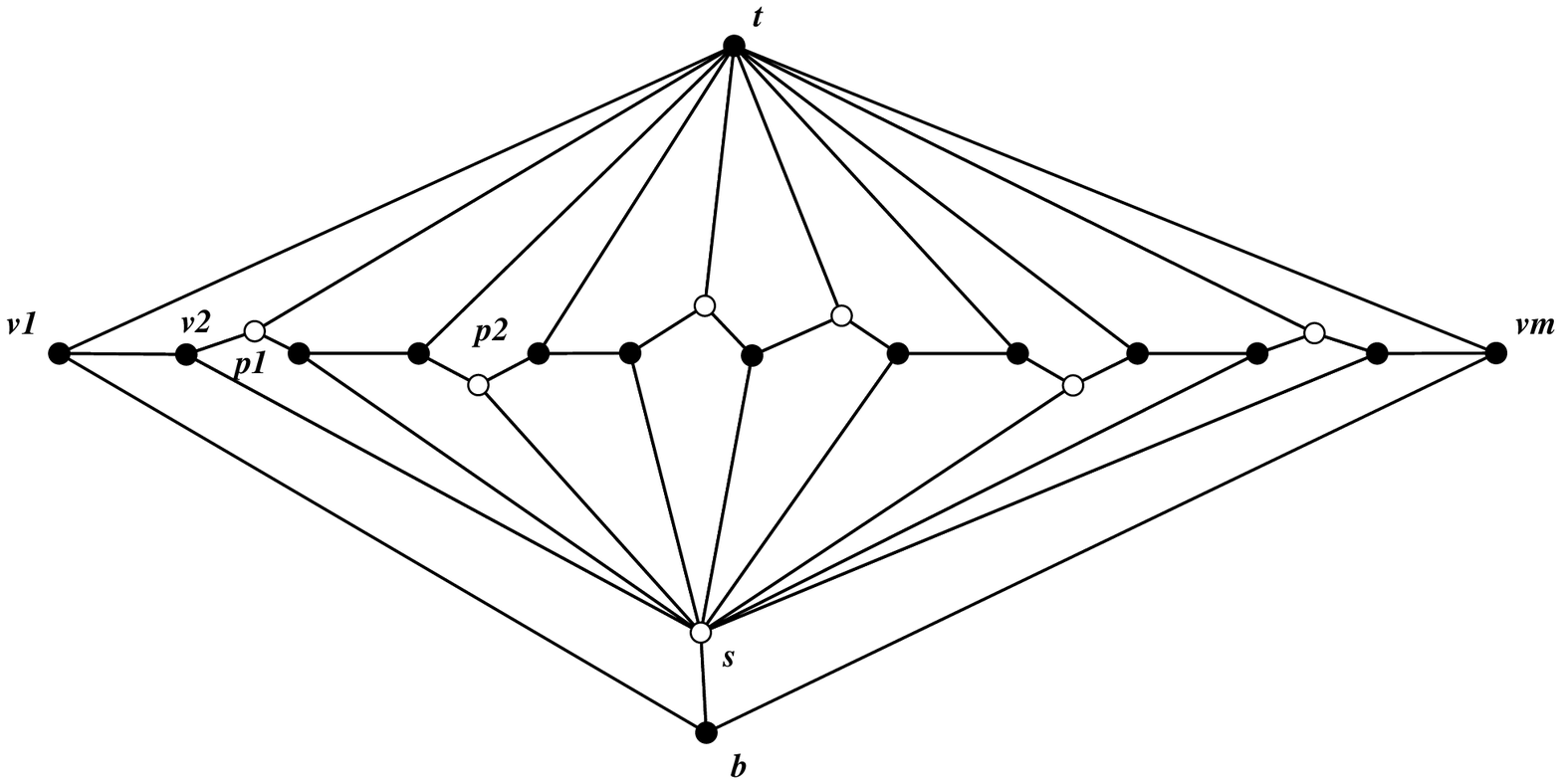}}
\Caption{A convex quadrangulation of $S$ using $\lceil{\frac{m}{2}}\rceil$ 
Steiner points.}
{fig:qline4}
\end{figure}

\begin{theorem}
The point set $S$ can be convex-quadrangulated with $s\leq \left \lceil\frac{ m+3}{2} \right \rceil$ Steiner points.
\end{theorem}

\begin{proof}
  \begin{shortonly}
    \emph{(Sketch)}
  \end{shortonly}
It is possible to convex-quadrangulate the given point set
configuration with $s$ Steiner points, where
\setlength{\extrarowheight}{4pt}
$$
s=
\left\{\begin{array}{ll}
{\frac{m}{2}}+1,    &\ \mathrm{if \ } m\equiv 0 \pmod{4}\\
{\frac{m+1}{2}}+1,  &\ \mathrm{if \ } m\equiv 1 \pmod{4}\\
{\frac{m}{2}}+2,    &\ \mathrm{if \ } m\equiv 2 \pmod{4}\\
{\frac{m+1}{2}},    &\ \mathrm{if \ } m\equiv 3 \pmod{4}
\end{array}\right\}\leq \left \lceil{\frac{m+3}{2}}\right\rceil
$$
A solution is presented in Figure \ref{fig:qline4}%
\begin{longonly}
, where the original
points are shown in black and the Steiner points in white%
\end{longonly}.
This solution can be described as follows.  Let $v_i$,
$i\in\{1,\dots,m+1\}$ be the points on the line $\ell$, and $t$ and
$b$ the top and bottom points. Place one Steiner point $s$ below
$\ell$, 
\begin{longonly}
inside the convex hull
\end{longonly}
 and in $L(b,v_2) \cap R(b,v_m)$.
 \begin{longonly}
 Quadrangles $\mathit{bsv_2v_1}$ and
$\mathit{bv_{m+1}v_ms}$, both of which are strictly convex, are part
of the quadrangulation.
 \end{longonly}
We call the line segment $v_iv_{i+1}$ (not necessarily part of the
quadrangulation)  the $i$th \emph{virtual edge} $e_i$.  
Suppose $m=4k+r$, $0\leq r \leq3$. Starting from both ends of $l$,
$2k$ Steiner points $p_i$ are placed alternately above and below every
other virtual edge on $\ell$.%
\begin{longonly}
More precisely,   for $1\leq i\leq
k$ place a Steiner point $p_i$ above (resp.\ below) $e_{2i}$ if $i$ is odd
(resp.\ even).  In both cases ensure $p_i$ is in the intersection of
the wedges $v_{2i}tv_{2i+1}$ and $v_{2i}sv_{2i+1}$. 
Connect $p_i$ to $t$ ($i$ odd) or $s$ ($i$ even), and to $v_{2i}$ and
$v_{2i+1}$. Connect $v_{2i-1}$ to $v_{2i}$. Carry out the
analogous procedure starting with the rightmost virtual edge.
\end{longonly}
After placing $2k$ Steiner points, we are left with
$r$ ``untreated'' virtual edges $e'_1, e'_2, \dots e'_r$ in the
center.  If $r\leq 2$, we place Steiner points as follows: one point
above (resp.\ below) each $e'_i$ if $k$ is odd (resp. even).  If $r=3$
then we place point below (resp.\ above) $e'_2$ if $k$ is odd (resp.
even). In all cases we insure the the Steiner point is within the two
wedges defined by the virtual edge, $s$ and $t$.
The strict convexity of the quadrangles created by this procedure is ensured 
by placing each Steiner point in the intersection of these two wedges.
\begin{longonly}
\begin{figure}[tbhp]                  
\psfrag{m4k}{$m=4k$}                                      
\psfrag{m4k1}{$m=4k+1$}                                      
\psfrag{m4k2}{$m=4k+2$}                                      
\psfrag{m4k3}{$m=4k+3$}                                      
\centerline{\includegraphics[width=.9\textwidth]{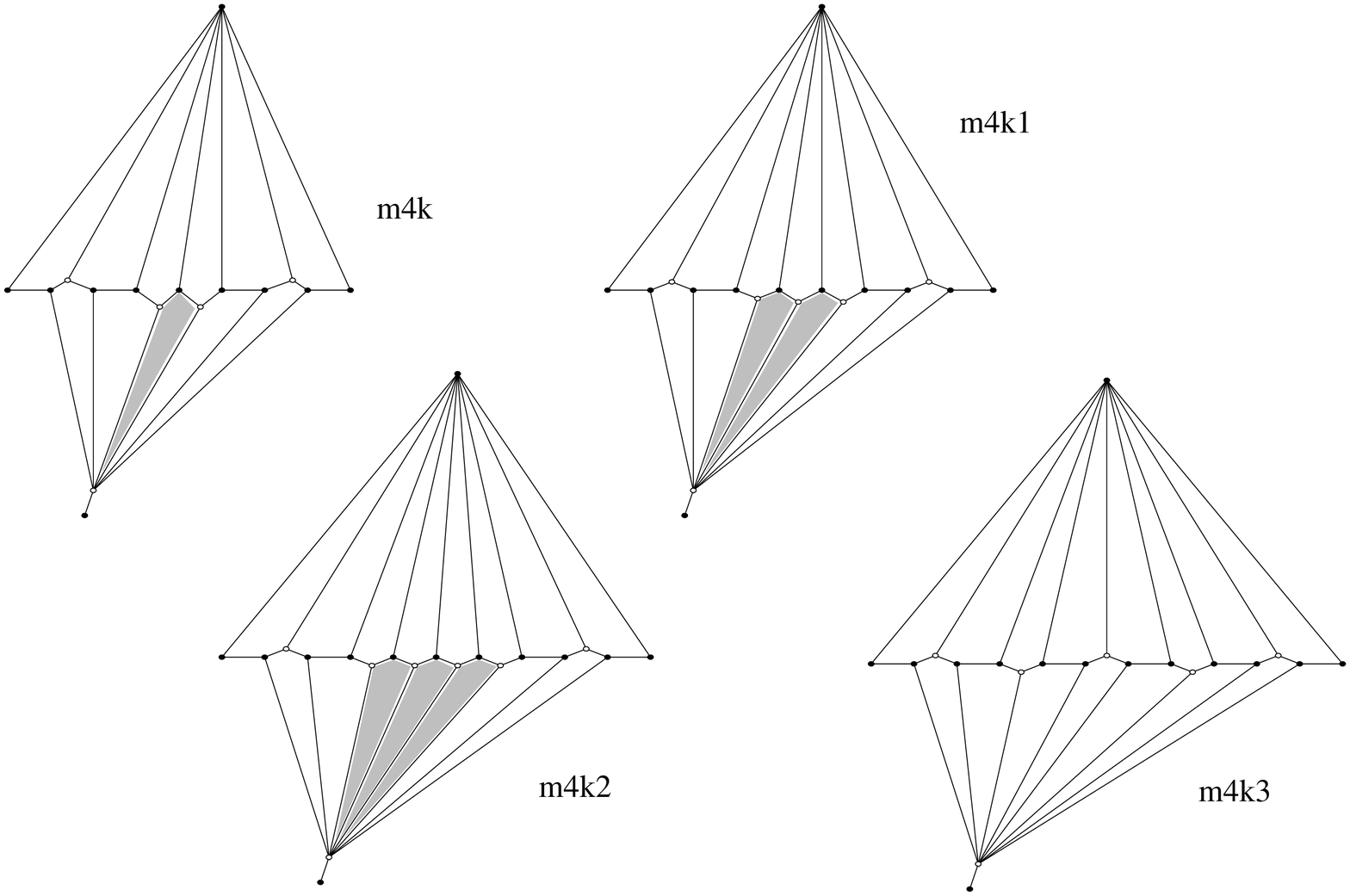}}
\caption{Convex quadrangulations for any parity of $m$ (shaded faces
are extra faces).}
\label{fig:qline5}
\end{figure}
The fact that the number of Steiner points used in these
quadrangulations is off by a small constant from the bound given by
our charging scheme is explained by the charges on the extra faces
(drawn shaded in Figure~\ref{fig:qline5}). In these cases, the extra
faces actually get charged more than $1$, whereas we count a charge of
only $1$ for any extra face of the quadrangulation when proving the
lower bound.
\end{longonly}
\qed\end{proof}

\begin{figure}[tbhp]
  \psfrag{ai}{$a_i$}
  \psfrag{bi}{$b_i$}
  \psfrag{ai1}{$a_{i+1}$}
\centerline{\includegraphics[width=0.4\columnwidth]{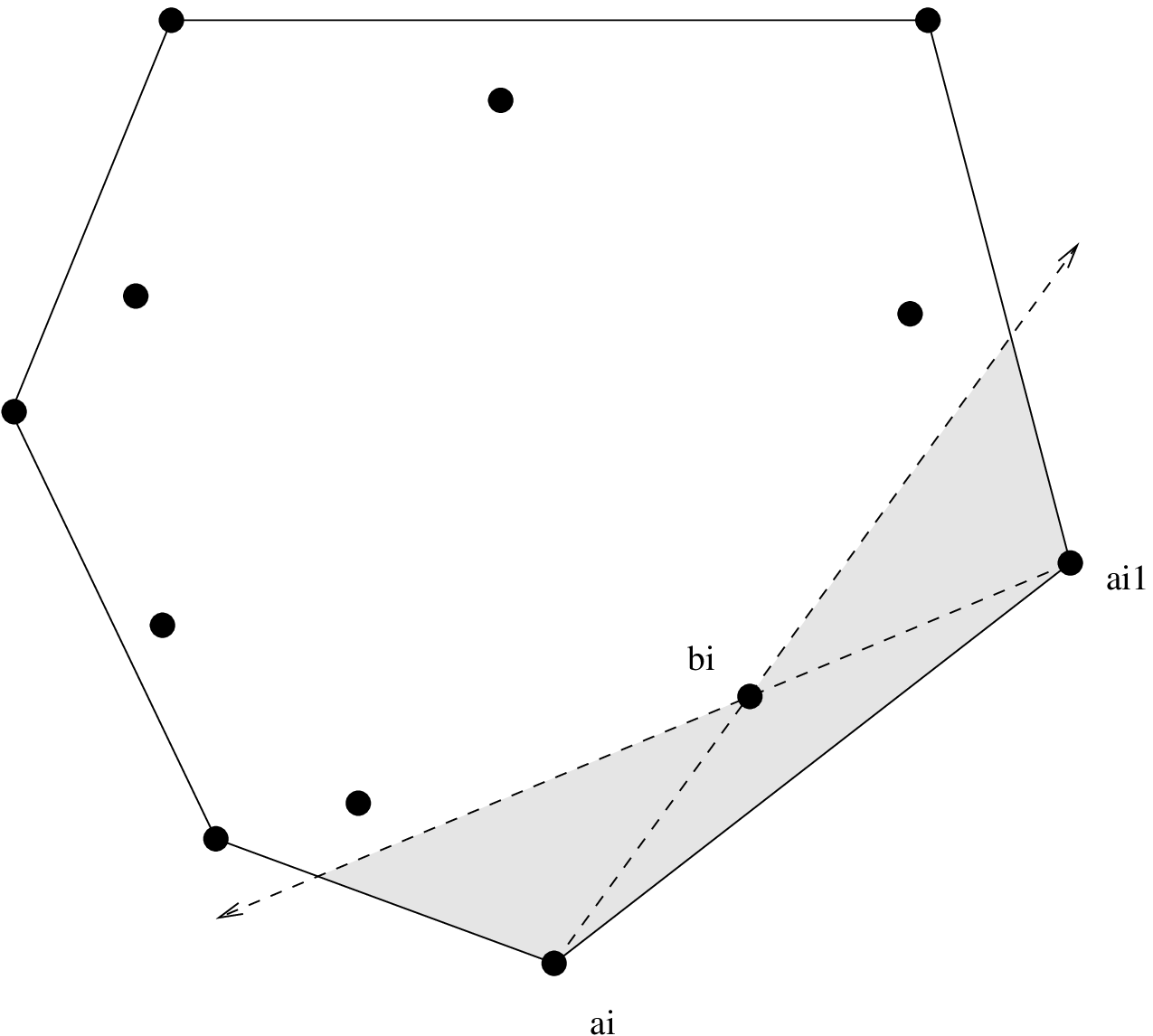}}
\caption{The point set $P$ has $n/2$ points on the convex hull, and $n/2$
interior points lying very close to edges of the convex hull.
}
\label{fig:qcircle1}
\end{figure}

Theorem~\ref{theorem:line} uses a highly degenerate
configuration, where most of the points lie on a
straight line. 
It turns out that the same lower bound result cannot be obtained
from this point configuration if it is perturbed%
\begin{longonly}%
: if the points do not
lie on a straight line, then squares can be formed using only input
points (i.e., without using Steiner points), and so can
diamonds.  
 It turns out that there exist analogous configurations 
 with  $m+1$ points on an arbitrary upward convex curve
(instead of a straight line), where the point set can be
convex-quadrangulated with a constant number of Steiner points%
\end{longonly}%
.
 We now describe a perturbable
(i.e.~non-degenerate) point set configuration that requires at
 least~$\frac{n}{4}$ Steiner points for a strictly convex quadrangulation. 

\paragraph{Description of the perturbable configuration of points:}
Let $n=2k$. Place $k$ points in convex position. Place the remaining
$k$ points such that, for each edge $e$ of the convex hull, there is one
point lying in the interior of the convex hull, very close to the midpoint 
of $e$. To be more precise, 
 if $(a_i,a_{i+1})$ is an edge of the
convex hull, the new point $b_i$ must be located so that $a_{i+2}\in
L(a_i,b_i)$ and $a_{i-1}\in R(a_{i+1},b_i)$, as illustrated in Figure
\ref{fig:qcircle1}. Call this point set $P$.

\begin{theorem}
The point set $P$ requires
at least $\frac{n}{4}$ Steiner points to be convex-quadrangulated.
\end{theorem}

\begin{proof}
By definition, each convex hull edge $(a_i,a_{i+1})$ must belong to one
quadrangle $Q_i$. For $Q_i$ to be convex and not contain any
interior point, its remaining two vertices must belong to the region
$G(i) = R(a_i,b_i)\cup L(a_{i+1},b_i)$; one of these vertices may be
$b_i$ (see Figure \ref{fig:qcircle1}).  
Hence, for every convex hull edge there is at least one Steiner point
in region $G(i)$. Since only consecutive regions intersect, at least
one Steiner point is needed for every pair of convex hull edges, i.e.\ at least $\frac{n}{4}$ Steiner points are needed.
\qed\end{proof}

\begin{figure}[tbhp]
\centerline{\includegraphics[width=0.4\columnwidth]{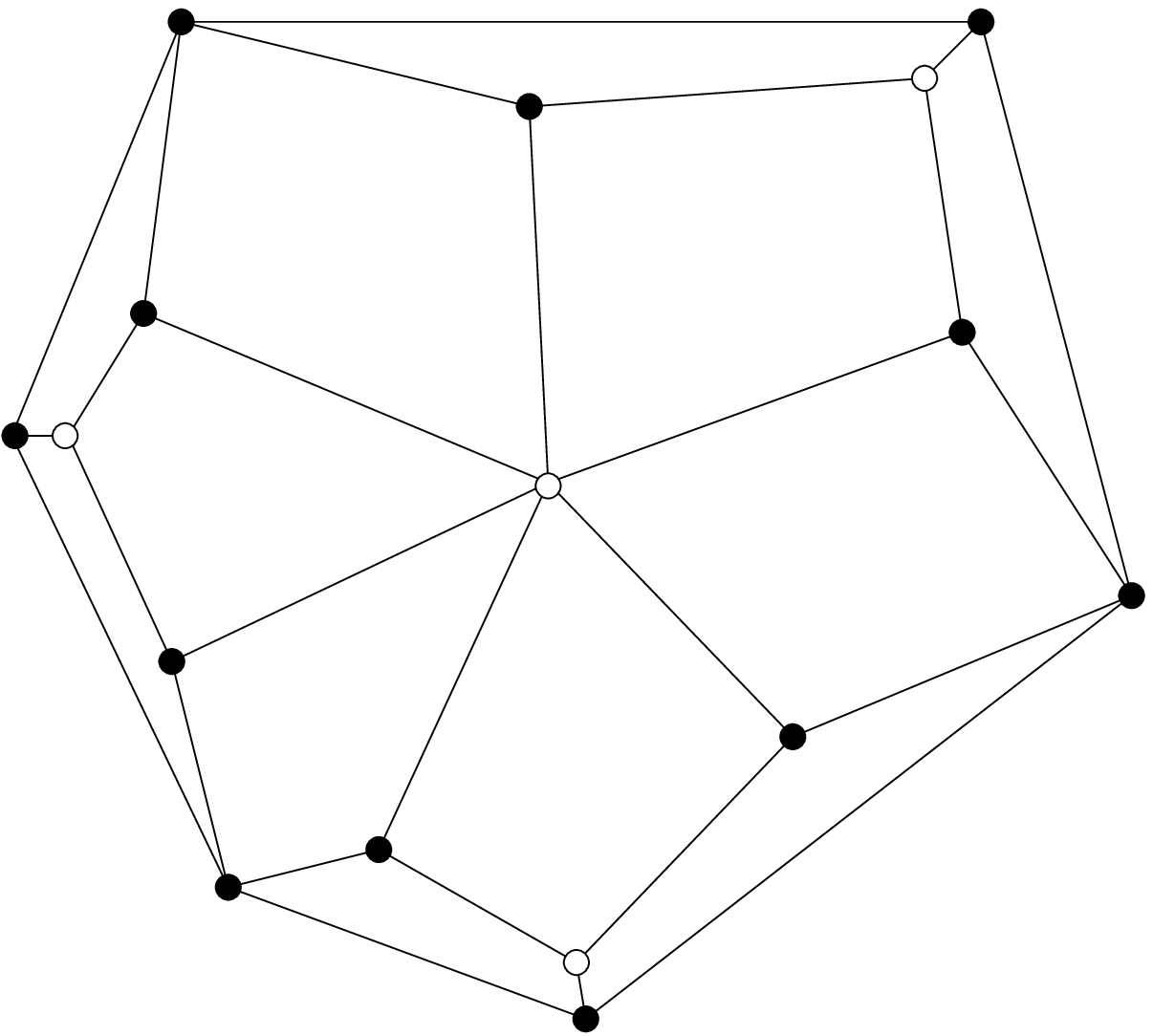}}
\caption{A convex-quadrangulation of $P$ that uses ${\frac{n}{4}}+1$
Steiner points.}
\label{fig:qcircle3}
\end{figure}

\begin{theorem}
 $P$ can be convex-quadrangulated with at most $\frac{n}{4} +1$ Steiner points.
\end{theorem}
  \begin{shortonly}
    Proof omitted; refer to Figure~\ref{fig:qcircle3}.    
  \end{shortonly}
  \begin{longonly}
\begin{proof}
    Figure \ref{fig:qcircle3} shows a convex-quadrangulation of the
    set that uses ${\frac{n}{4}}+1$ Steiner points. There is one
    quadrangle for every convex hull edge $a_ia_{i+1}$. It has $b_i$
    as a vertex and uses one Steiner point, which is shared by the
    adjacent convex hull edge.  Finally, one central Steiner point is
    used to convex-quadrangulate the remaining interior face.
    
    In this configuration $n$ is always even.  If $n\equiv 2
    \pmod{4}$, then we have an odd number of points on the convex
    hull, and $\lceil{\frac{n}{4}}\rceil$ Steiner points suffice to
    convex-quadrangulate with an extra triangle, formed by one of the
    convex hull edge and its corresponding interior point.
    \qed\end{proof}
  \end{longonly}

\hexlabel
\section{Upper bound\label{sec:upper}}

Given a set $S$ of $n$ points in the plane, $\CH(S)$ is the convex 
hull of $S$. For a simple polygon $P$, $\interior(P)$ denotes the 
interior of $P$ and $\kernel(P)$ is the locus of points in $P$ 
that can see all of $P$.
A {\em
convex quadrangulation} of $S$ is a decomposition of $\CH(S)$ into
strictly convex quadrangles and at most one triangle, such that no cell 
contains a point of $S$ in its interior. The vertices of
the quadrangulation that do not belong to $S$ are called {\em Steiner
points}. In what follows, 
angles greater than or equal to $180^\circ$ are {\em reflex angles}.

\begin{figure}[tbph]
  \begin{center}
    \includegraphics[width=\figwidth]{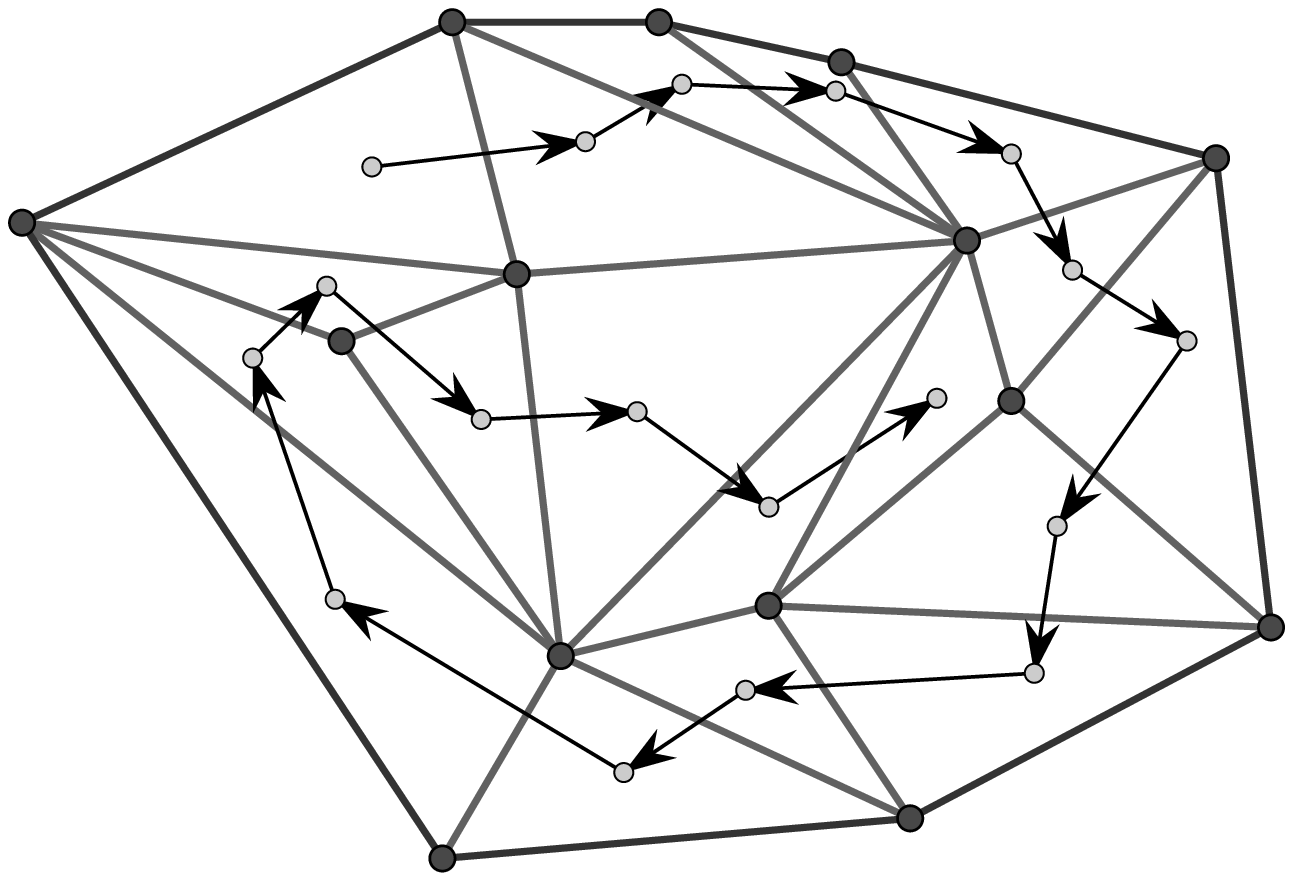}
    \caption{A path triangulation of a point set}
    \label{fig:pathtri}
  \end{center}
\end{figure}

\begin{theorem}
Any set of $n$ points can be convex-quadrangulated using at most
$3{\lfloor{\frac{n}{2}}\rfloor}$ Steiner points.
\end{theorem}

\begin{proof}
  \begin{shortonly}
    \emph{(Sketch)}
  \end{shortonly}
Any set $S$ of $n$ points has a path triangulation (a triangulation
whose dual graph has a Hamiltonian path), which can be constructed
in $O(n\log n)$ time~\cite{AHMS94,bt97} (Figure~\ref{fig:pathtri} 
illustrates such a triangulation of a point set).
Denote by $t$ the number of triangles in any triangulation of $n$
points with $h$ extreme points ($t=2n-2-h$).  By pairing up the
triangles along the path, we obtain a path quadrangulation of $S$ with
possibly one unpaired triangle (see Figure~\ref{fig:pairtri}).
\begin{figure}[tbhp]
  \begin{center}
    \includegraphics[width=\figwidth]{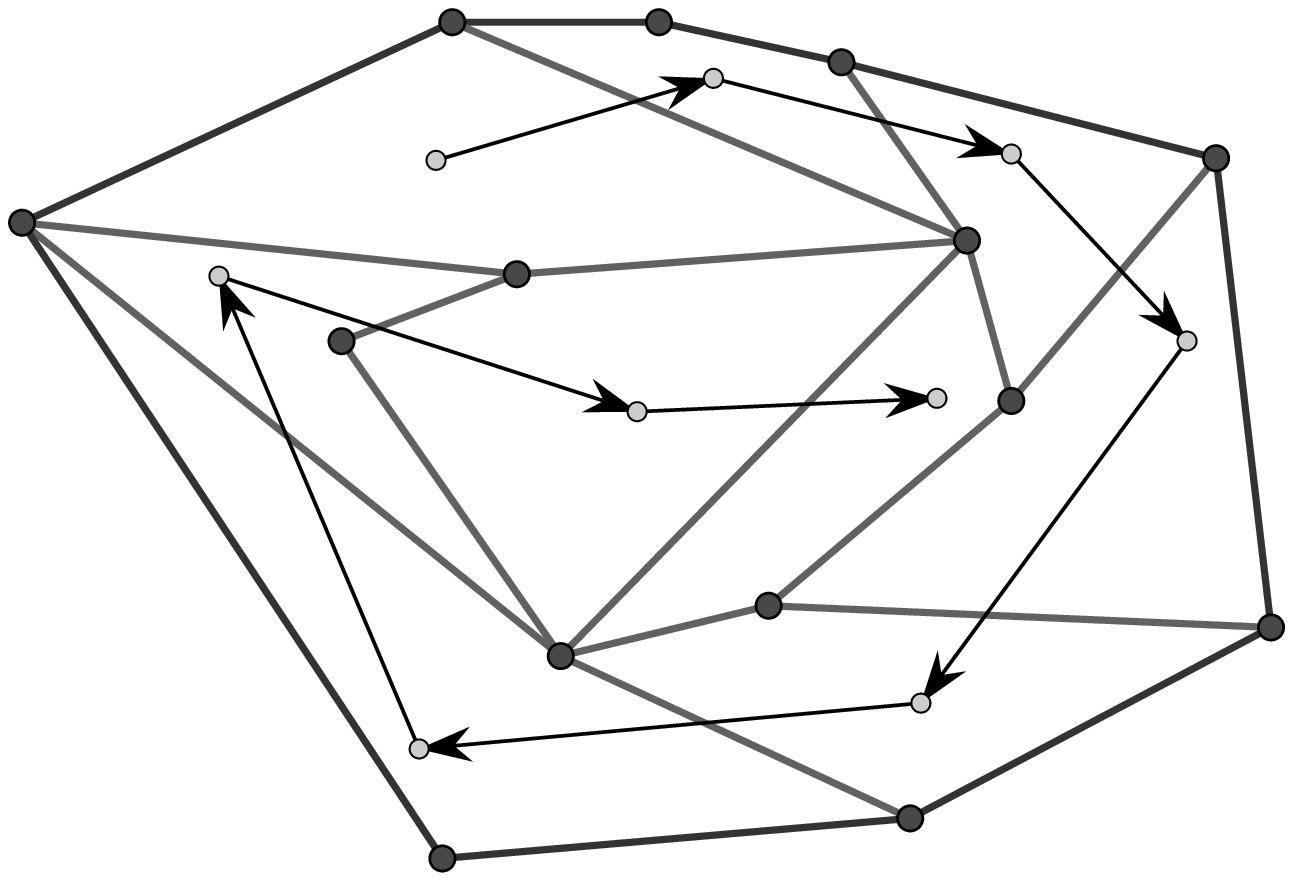}
    \caption{The triangles of Figure~\protect\ref{fig:pathtri} paired to form
      a (not necessarily convex) quadrangulation}
    \label{fig:pairtri}
  \end{center}
\end{figure}
 We will
prove in Section~\ref{sec:pairing} that it is always possible to
convex-quadrangulate a pair of consecutive quadrangles by using at
most 3 internal Steiner points.
\begin{longonly}
 At the end of
the process we may have any of the following situations:

\begin{itemize}
\item 
There is no unpaired triangle ({\em i.e.}, $h$ is even) and all the
quadrangles have been paired up.  In this case, a
convex-quadrangulation has been obtained with no leftover triangle.
Therefore, the number of quadrangles is $q={\frac{t}{2}}=n-1-{\frac{h}{2}}$,
and the total number of Steiner points used is $ s={\frac{3}{2}}q
={\frac{3}{2}}(n-1-{\frac{h}{2}})\leq 3\lfloor{\frac{n}{2}}\rfloor$.

\item 
There is one unpaired triangle ({\em i.e.}, $h$ is odd), all the
quadrangles have been paired up.  In this case, the number of
quadrangles is $q={\frac{t-1}{2}}$, which is less than in the previous
case. Once again, the number of Steiner points $s={\frac{3}{2}}q<
3\lfloor{\frac{n}{2}}\rfloor$.

\item 
There is no unpaired triangle and all the quadrangles except one have
been paired up. In this case, the last quadrangle can be convex-quadrangulated,
if it is not convex, by adding 4 internal Steiner points (see
page~\pageref{quadconv} for details). Since all quadrangles except one
have been paired up, the number of Steiner points used is
$s={\frac{3}{2}}(q-1)+4=
{\frac{3}{2}}({\frac{t}{2}}-1)+4={\frac{3}{2}}(n-2-{\frac{h}{2}})+4\leq  
{\frac{3}{2}}n-{\frac{3}{4}}h+1 < 3{\lfloor{\frac{n}{2}}\rfloor}$
since $h\geq 4$.

\item 
There is one unpaired triangle, and all the quadrangles except one
have been paired up.
We can convex-quadrangulate the remaining quadrangle with 4 Steiner
points as before, and leave the triangle as it is. In this case, we
have $q={\frac{t-1}{2}}$ and
$s={\frac{3}{2}}(q-1)+4<3\lfloor{\frac{n}{2}}\rfloor$ (as argued in
the previous case). 
\end{itemize}
\end{longonly}
\begin{shortonly}
  Consideration of the various possibilities for unpaired triangles
  and quadrangles yields the bound above.
\end{shortonly}
Note that the number of quadrilaterals in the quadrangulation is at
most $5{\lfloor{\frac{n}{2}}\rfloor} - {\frac{h}{2}}$. Note also that 
the quadrangulation produced by our algorithm is strictly convex, even 
if the path quadrangulation contains degenerate quadrilaterals.
\qed\end{proof}

\subsection{Pairing up quadrangles \label{sec:pairing}}

\renewcommand{\theenumii}{\arabic{enumii}}
\settowidth{\leftmarginii}{1.}
\settowidth{\leftmargini}{1. }
\renewcommand{\labelenumii}{\theenumi.\theenumii.}
\newcommand{\Case}[1]{\item ($\mathit{#1}$)\hspace{0.7em}}

Before discussing the details of how to convex-quadrangulate a pair
of adjacent quadrilaterals, we introduce some notation and mention
a few useful facts about polygons.
Given two points $p$ and $q$, we will denote by $L(p,q)$ (resp.\ 
$R(p,q)$) the left (resp. right) open half-plane defined by the
oriented line from $p$ to~$q$. Throughout this section, vertices of
polygons will be enumerated counterclockwise. Given a vertex $v$ of a
polygon $P$, we denote its successor (resp. predecessor) by $v^+$
(resp.\ $v^-$), and we write $\Wedge(v)$ to mean $L(v^-,v) \cap R(v^+,v)\cap
\interior(P)$.  If $v$ is reflex, $\Wedge(v)$ will denote  
the locus of points (inside $P$) that can be connected to $v$ forming
strictly convex angles at $v$. If $v$ is convex, $\Wedge(v)$ is the
interior of the visibility region of $v$ in $P$. Given three points
$p, q,$ and $r$, $\triangle(pqr)$ is the open triangle defined by the
three points, i.e.~$\triangle(pqr)=\interior\CH(p,q,r)$. 
 Note that
 \begin{equation}
   \label{eq:int-kern}
\interior\kernel(P) ={\cap}_{v\in P}L(v,v^+)\,.  
 \end{equation}
\begin{longonly}
We can observe the
following
  \begin{align*}
    \Wedge(v_i)&=L(v_i^-,v_i) \cap R(v_i^+,v_i)\\
    &=L(v_i^-,v_i) \cap L(v_i,v_i^+)
  \end{align*}
It follows that 
\begin{equation}
  \label{eq:kernwedge}
    \interior \kernel(P)=\bigcap_{i} \Wedge(v_{2i})\,.
\end{equation}
Similarly, by noting that if $v_0 \dots v_k$ form a reflex chain,
\begin{equation*}
  \bigcap_{0\leq i <k} L(v_i,v_{i+1}) = L(v_0,v_1) \cap L(v_{k-1},v_k)\,,
\end{equation*}
it follows that 
\begin{equation}
  \interior \kernel (P) = 
\bigcap_{v \text{ convex}} \Wedge(v)\,.
  \label{eq:kernconv}
\end{equation}
\end{longonly}
\begin{shortonly}
The following observations will prove useful below.  
\begin{align}
  \label{eq:kernwedge}
    \interior \kernel(P)&=\bigcap_{i} \Wedge(v_{2i})\\
&=\bigcap_{\squash{\ensuremath{v}\text{ convex}}} \Wedge(v)
  \label{eq:kernconv}
\end{align}
\end{shortonly}

\begin{figure}[tbhp]
  \begin{center}
    \includegraphics[width=\figwidth]{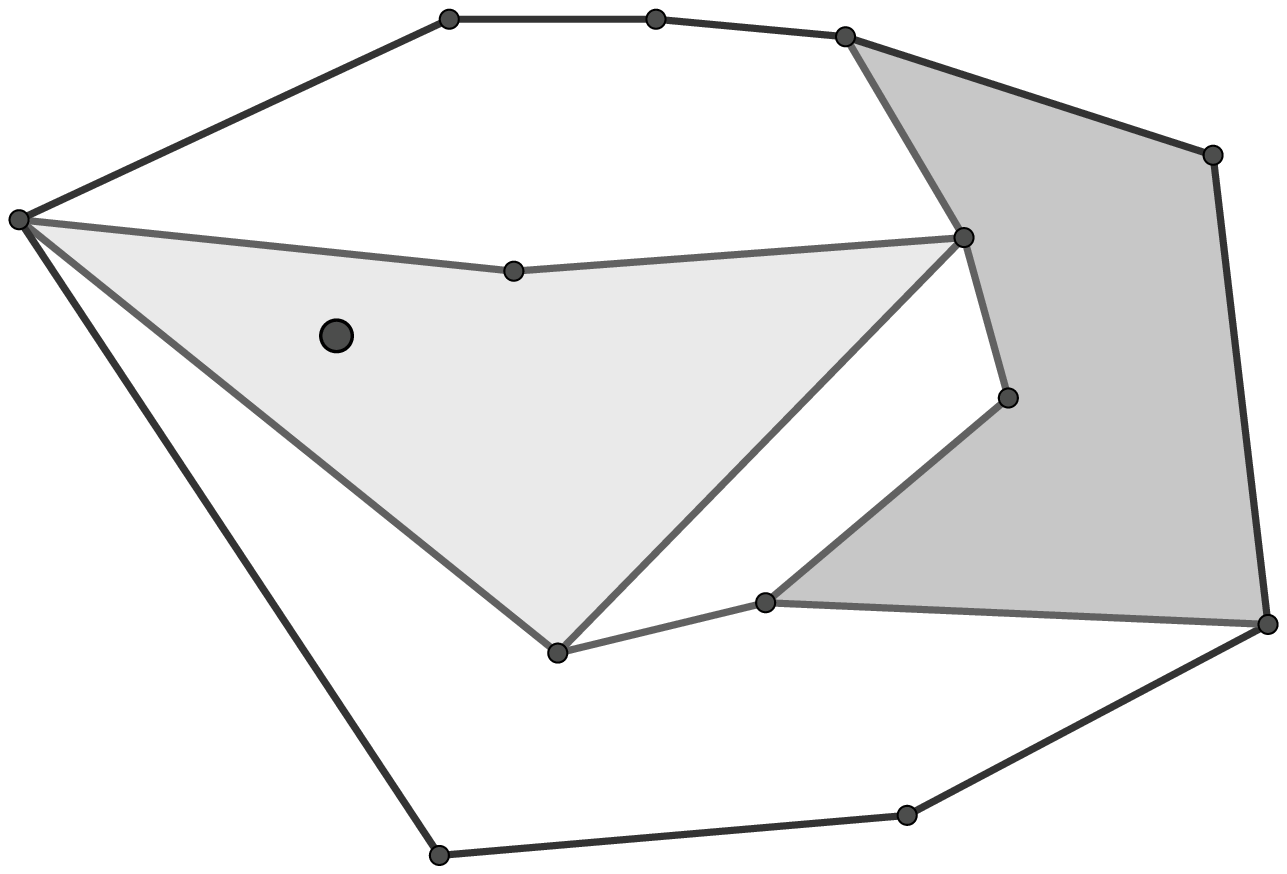}
    \caption{The union of a pair of adjacent quadrangles is either
      a hexagon or quadrangle with a fifth interior point. }
    \label{fig:pairquad}
  \end{center}
\end{figure}

\begin{figure}[tbhp]
  \begin{center}
    \includegraphics[width=\figwidth]{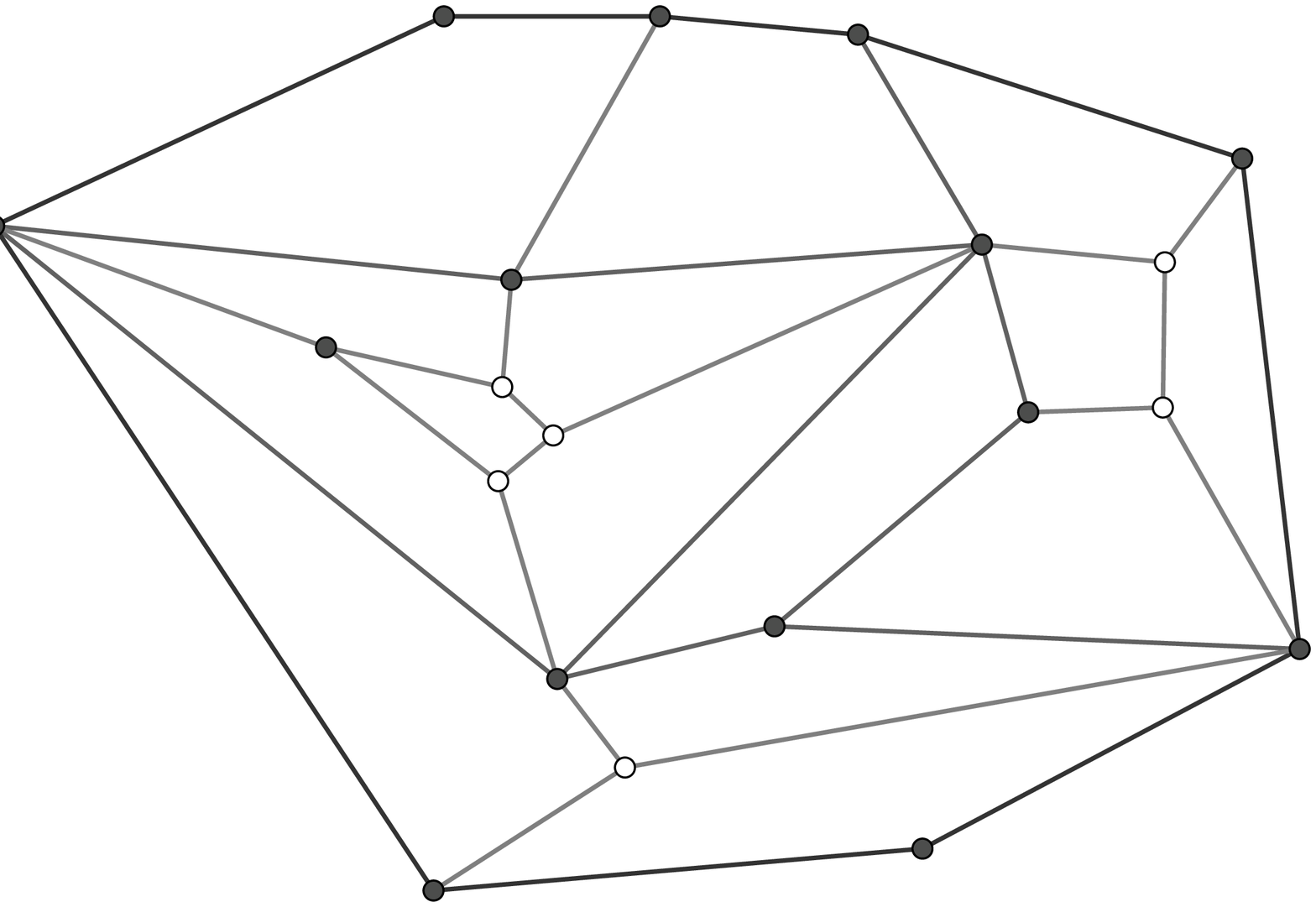}
    \caption{A convex
      quadrangulation resulting from applying our techniques to the
      point set of Figures~\ref{fig:pathtri}, \ref{fig:pairtri}, 
      and \ref{fig:pairquad}. }
    \label{fig:quadex}
  \end{center}
\end{figure}

Consider a pair of consecutive quadrangles in the path
quadrangulation. They may share one edge or two edges. In the first
case, their union is a hexagon, while in the second case it is a
quadrangle containing a fifth point in its interior (see
Figure~\ref{fig:pairquad}).
  In the rest of
this section we will examine in detail how to convex-quadrangulate the
union of two quadrangles.
\begin{longonly}
  The general scheme will be inductive, i.e.\ 
to reduce each case to one requiring fewer Steiner points by the
addition of a single Steiner point.  
\end{longonly}
\begin{table}[tbhp]                                                        
  \begin{center}
{\includegraphics[width=.8\textwidth]{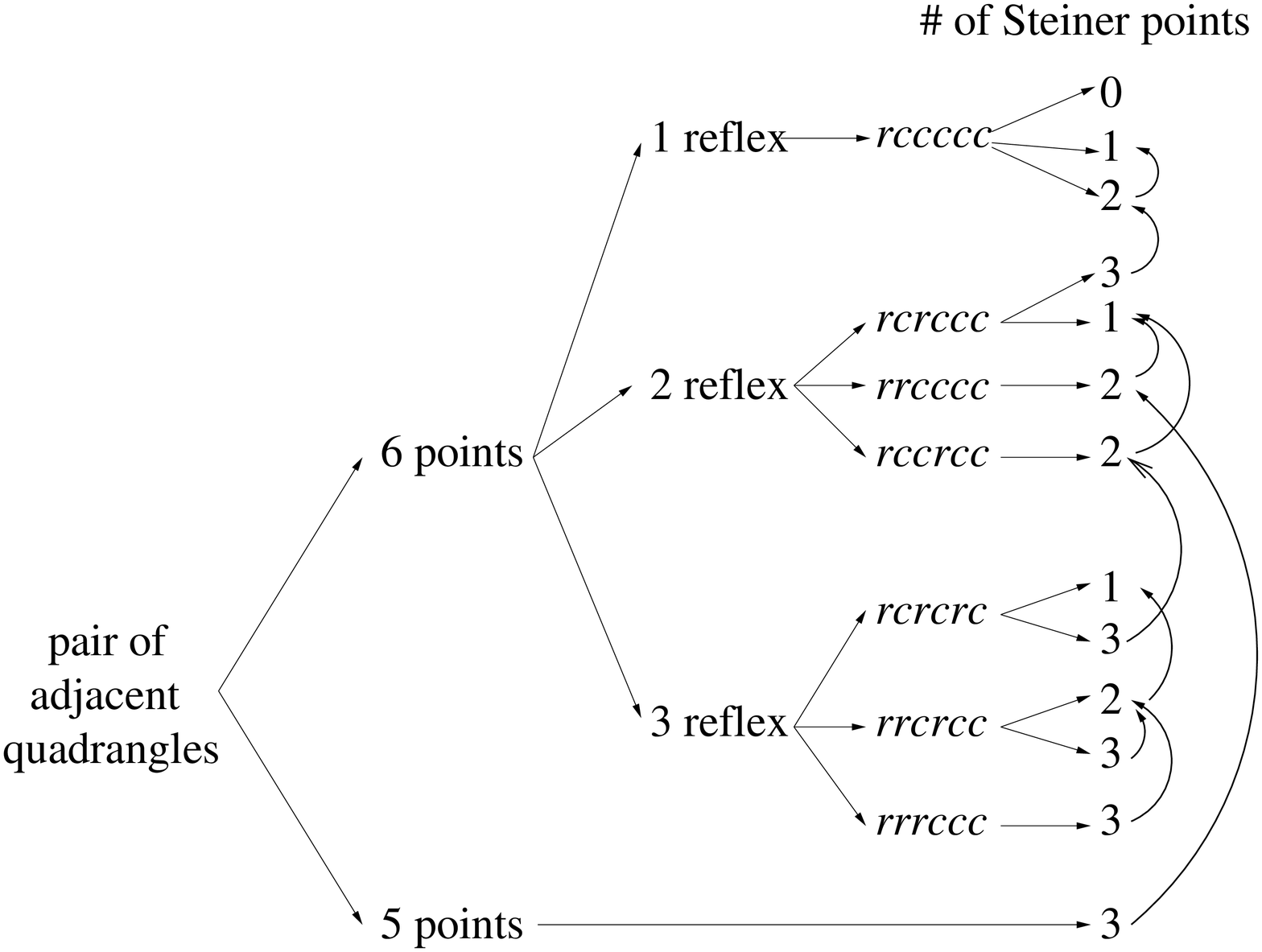}}
\caption{Scheme of the proof.}
\label{table:tree} 
  \end{center}
\end{table}
Table \ref{table:tree} provides a
summary of all the cases and their interdependencies. Most of the
cases are given a mnemonic label describing the cyclic order of reflex
and convex vertices around the polygon boundary and the total number
of Steiner points necessary (e.g.\ $rcrcrc$-1 describes the case
where reflex and convex vertices alternate and one Steiner point
suffices to convex-quadrangulate the hexagon).  The last column
reports the number of Steiner points used in each case.  The arrows on
the right indicate the reductions, after adding one Steiner point,
from one case to another.
As is suggested by Table~\ref{table:tree}, the majority of our effort
in the remainder of this section will be devoted to proving the
following theorem.
\begin{theorem}
  \label{theorem:hexquad}
  Any hexagon can be convex-quadrangulated by placing at most $3$
  Steiner points in its interior.
\end{theorem}
The curious reader is referred to Figure~\ref{fig:quadex} for a convex
quadrangulation resulting from applying our techniques to the point
set of  Figures~\ref{fig:pathtri}, \ref{fig:pairtri}, 
      and~\ref{fig:pairquad}.  In the figure the white points are Steiner points.

\subsubsection{Independent Triples}

We call a set of vertices of a polygon \emph{independent} if no two of
them are endpoints of the same edge.  
\begin{longonly}
We start by establishing some
useful properties of independent triples of vertices of a hexagon.
\end{longonly}
All lemmas in this section hold even when reflex angles are exactly 
equal to $180^\circ$. 
Let $\set{a,c,e}$ be an independent triple for
a hexagon $P=abcdef$.

\begin{lemma}\label{lemma:alfa}
  If
  $\triangle(ace)\subset P$ then 
  $\triangle(ace)\cap\Wedge(a)=\triangle(ac'e')$, where $c'e'\subseteq ce$ and
  $c'\not=e'$.
\end{lemma}

\begin{longonly}
\begin{proof}
  It suffices to establish that   $ce \cap \Wedge(a)$ is a non-trivial
  line segment. The result then follows by convexity. 
  If $a$ is convex, then $c$ and $e$ are in the visibility polygon of
  $a$, i.e.\ in $\Wedge(a)$.  Suppose then that $a$ is reflex.
  If one of $c$ or $e$ is contained in $\Wedge(a)$ then the lemma holds.
 If neither $c$ nor $e$ belongs to
$\Wedge(a)$, then they cannot both belong to $R(a^-,a)$
(resp.\ $L(a^+,a)$) because then $a^-$ (resp.\ $a^+$) cannot see $c$ or $e$, which is a
contradiction because $a^{--}$ (resp.\ $a^{++}$) must be $c$ or $e$, since $a$,
$c$ and $e$ are at distance two. Therefore, $c$ and $e$ must be on
opposite sides of $\Wedge(a)$ ({\em i.e.}, one in $R(a^-,a)$
and the other in $L(a^+,a)$), and the segment $ce$ must have non-trivial
intersection with $\Wedge(a)$.
\qed\end{proof}
\end{longonly}

\begin{lemma}
  \label{lemma:twowedge}
If
  $\triangle(ace)\subset P$
  then $\Wedge(a) \cap \Wedge(c)\cap  \triangle(ace) \not = \emptyset$.
\end{lemma}

\begin{longonly}
\begin{proof}
  This follows by applying Lemma~\ref{lemma:alfa} twice, and convexity.
\qed\end{proof}
\end{longonly}
\begin{lemma}\label{lemma:beta}
If $\triangle(ace) \subset P$ then 
$\triangle(ace)\cap \Wedge(a^-)\cap
\Wedge(a^+)\not=\emptyset$.  
\end{lemma}

\begin{longonly}
\begin{proof}
  Note that $a^+$ and $a^-$ must both be convex (or exactly $180^\circ$). It follows that in a
  neighborhood $N(a)$ of the point $a$, we have $N(a)\cap
  \Wedge(a)=N(a)\cap \Wedge(a^-)\cap
  \Wedge(a^+)$, and Lemma \ref{lemma:alfa} applies.
\qed\end{proof}
\end{longonly}

\begin{lemma}
  \label{lemma:star}
  If $P$ is starshaped and $\triangle(ace) \subset P$,
  then one Steiner point suffices to convex-quadrangulate $P$.
\end{lemma}

\begin{longonly}
\begin{proof}
  From \eqref{eq:kernwedge}, $\interior \kernel(P)= \Wedge(a)\cap
  \Wedge(c)\cap \Wedge(e)$.  Each pair of these wedges intersect
  $\triangle(ace)$, as a consequence of Lemma~\ref{lemma:twowedge}.
  Each pair of wedges intersects as a consequence of
  Lemma~\ref{lemma:beta}. In this case we consider the wedges extended
  to the entire plane, and not restricted to the polygon.  Since both
  the triangle and the (extended) wedges are convex, Helly's
  theorem~\cite{w-httgt-97} applies. It follows they all intersect, i.e.\ the
  triangle $\triangle(ace)$ must intersect the interior of the kernel.

  One Steiner point $s$ can then be placed in the intersection of the
  triangle and the kernel, and connected to the three reflex vertices
  (see Figure \ref{fig:q63_1}). Since $s$ belongs to the kernel, it
  belongs to the wedges of the three reflex vertices, hence $a, c,$
  and $e$ are now strictly convex vertices in the quadrangulation. 
  Since $s$ belongs to
  $\triangle(ace)$, $s$ is convex in all the quadrangles.
  \qed\end{proof}
\end{longonly}

\begin{longonly}
    \begin{figure}[tbhp]  

          \centerline{\includegraphics[width=.5\columnwidth]{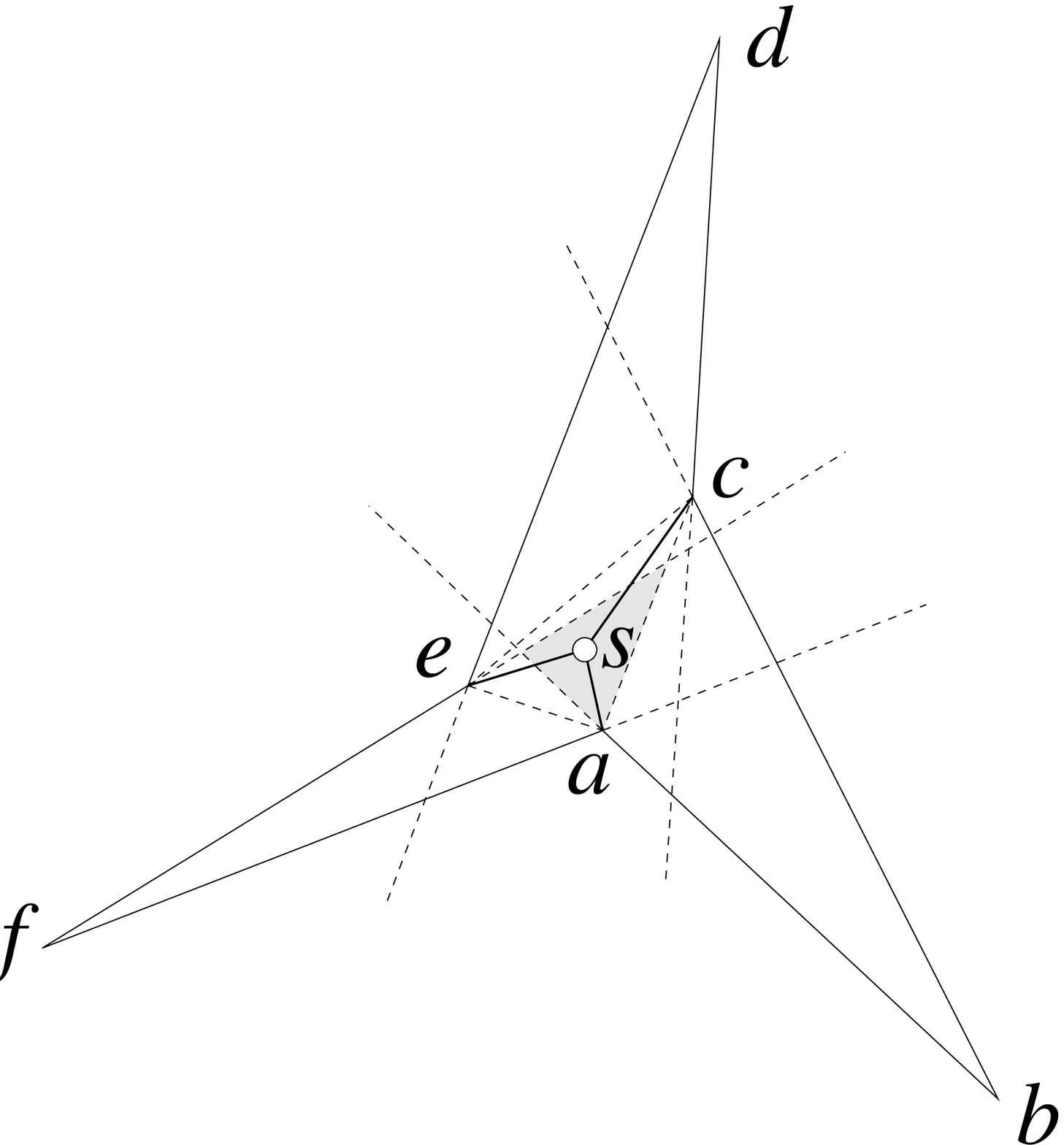}}

      \caption{One Steiner point suffices if $\triangle(abc)\subset P$ and $P$ is starshaped.}
      \label{fig:q63_1}

\end{figure}
\end{longonly}    

\begin{figure}
    \begin{center}
      \psfrag{a}{$a$}
      \psfrag{b}{$b$}
      \psfrag{c}{$c$}
      \psfrag{d}{$d$}
      \psfrag{e}{$e$}
      \psfrag{f}{$f$}
      \includegraphics[height=3in]{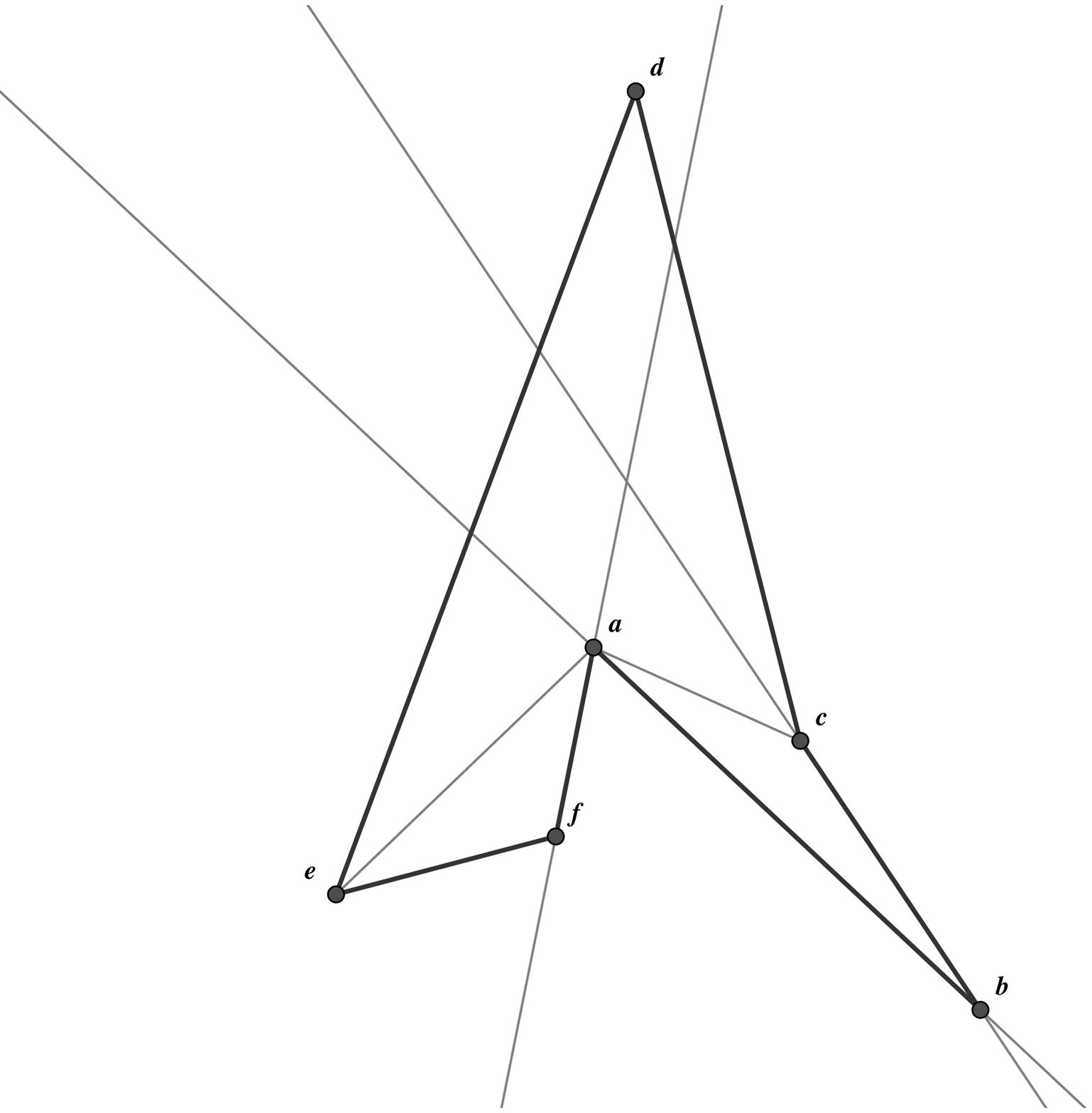}
    \caption{Illustrating the proof of Lemma~\protect\ref{lemma:obstruct}}
    \label{fig:obstruct}
  \end{center}
\end{figure}

\begin{lemma}
  \label{lemma:obstruct}
  If $c$ does not see $e$, and $a$ is the only reflex vertex other than 
  possibly $c$ or $e$, then
  \begin{thrmlist}
    \item $\Wedge(a) \cap \Wedge(c) \not=\emptyset$, and
    \item $\Wedge(a) \subset L(a,c)$.
  \end{thrmlist}
\end{lemma}

\begin{proof}\mbox{}
  \begin{thrmlist}
  \item Since $b$ and $f$ are convex, $a \in \triangle(cde)$, since
    otherwise nothing can block $ce$ (notice that both $abc$ and $aef$
    are ears of the polygon).  Similarly, the (non-convex)
    quadrangle $acde$ must be empty (see Figure~\ref{fig:obstruct}). It
      follows that $a$ sees $\set{c,d,e}$.  We can conclude that $a\in
      R(d,c)$ (by seeing $d$) and $a \in L(b,c)$ (by seeing $c$ and
      $d$). In other words, $a \in \Wedge(c)$.  The claim then follows
      from the fact that for some neighborhood $N(a)$, $N(a) \cap
      \Wedge(a) \subset \Wedge(c)$.
  \item Since $a \in \triangle(cde)$, it follows that $e \in R(a,c)$.
    Again considering the fact that $aef$ forms an ear of the polygon, 
    we have $f \in R(a,c)$.  It follows that both of the chords defining 
    $\Wedge(a)$ are contained in $L(a,c)$.
  \end{thrmlist}
\qed\end{proof}

\subsubsection{Proof of Theorem~\protect\ref{theorem:hexquad}.}

\begin{longonly}
We are now ready to carry out the case analysis described in
Table~\ref{table:tree}.  
\end{longonly}
A hexagon may have zero, one, two, or three reflex vertices; we
consider each of these cases in turn. 
In the remainder of this section, we will use {\em convex} to mean 
{\em strictly convex}.

\paragraph{Hexagon with no reflex vertices.}
In this case, the hexagon can be trivially decomposed into two convex
quadrangles without using any Steiner points.

\paragraph{Hexagon with one reflex vertex.}
Suppose w.l.o.g. that vertex $a$ is reflex. 
\begin{enumerate}
\Case{rccccc{-}0}
If $d\in \Wedge(a)$ then no Steiner points are needed.
Connecting $d$ with $a$ will produce a convex quadrangulation of the
hexagon, as shown in Figure \ref{fig:q61_1}. Note that if vertex $a$ 
is equal to $180^\circ$, this case must be satisfied.

\begin{longonly}
  \begin{figure}[tbhp]
    \centerline{\includegraphics[width=0.4\columnwidth]{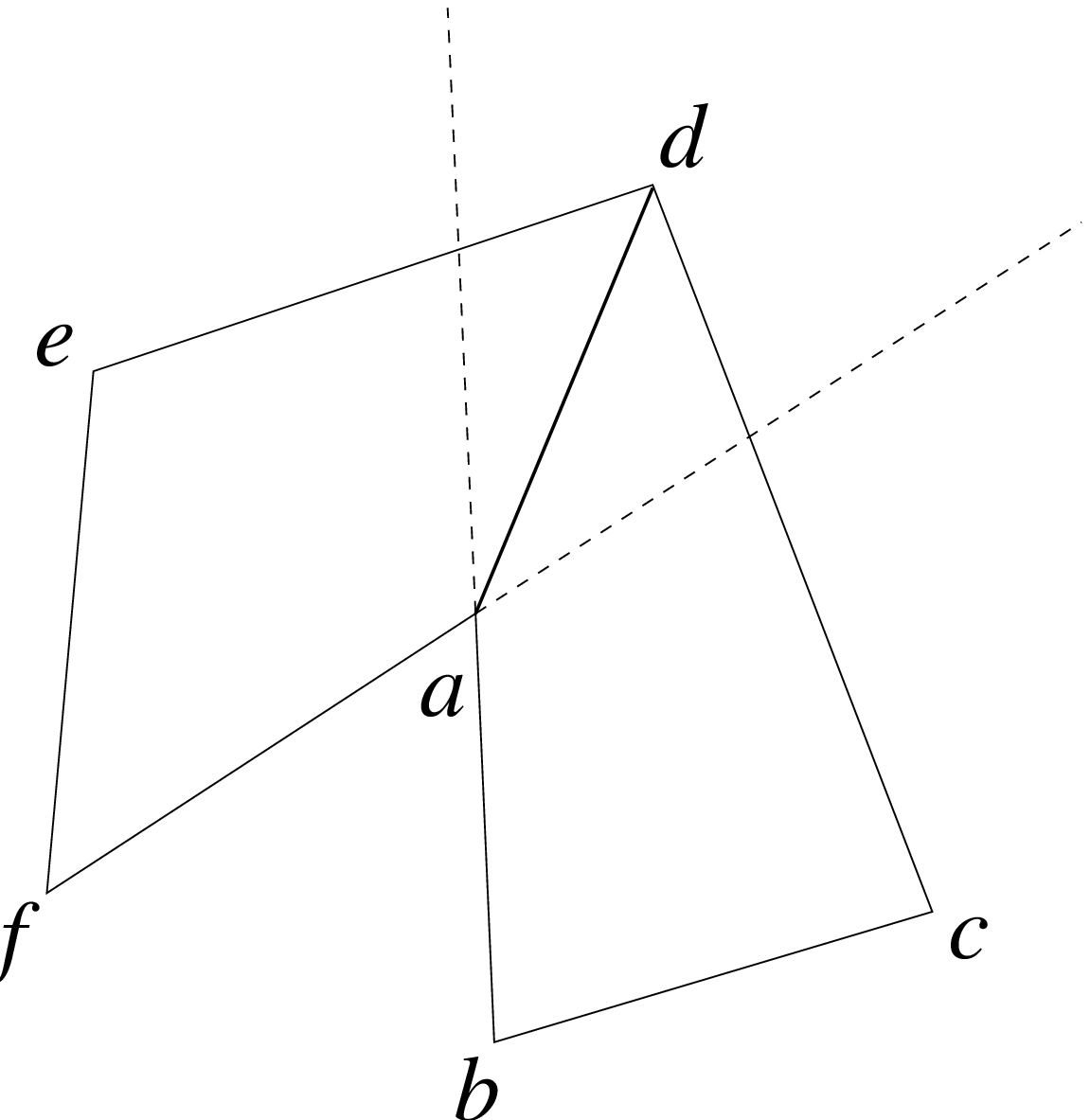}}
    \caption{No Steiner points needed.}
    \label{fig:q61_1}
  \end{figure}
\end{longonly}
\begin{figure}[tbhp]                                                        

    \centerline{\includegraphics[width=0.4\columnwidth]{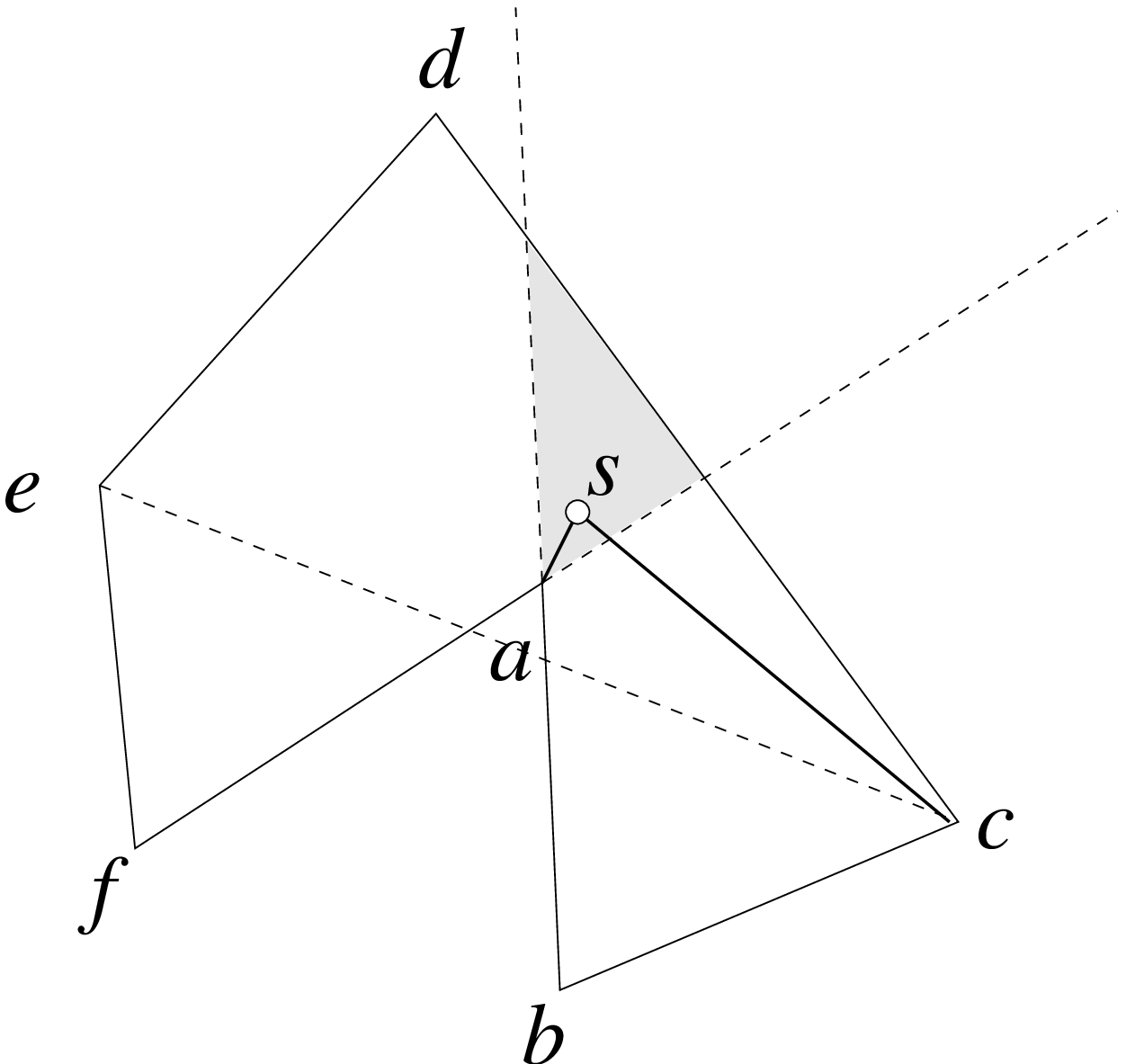}}
    \caption{One Steiner point reduces case $rccccc$-2 to case $rccccc$-1.}
    \label{fig:q61_3}

\end{figure}
\begin{figure}

      \centerline{\includegraphics[width=.4\columnwidth]{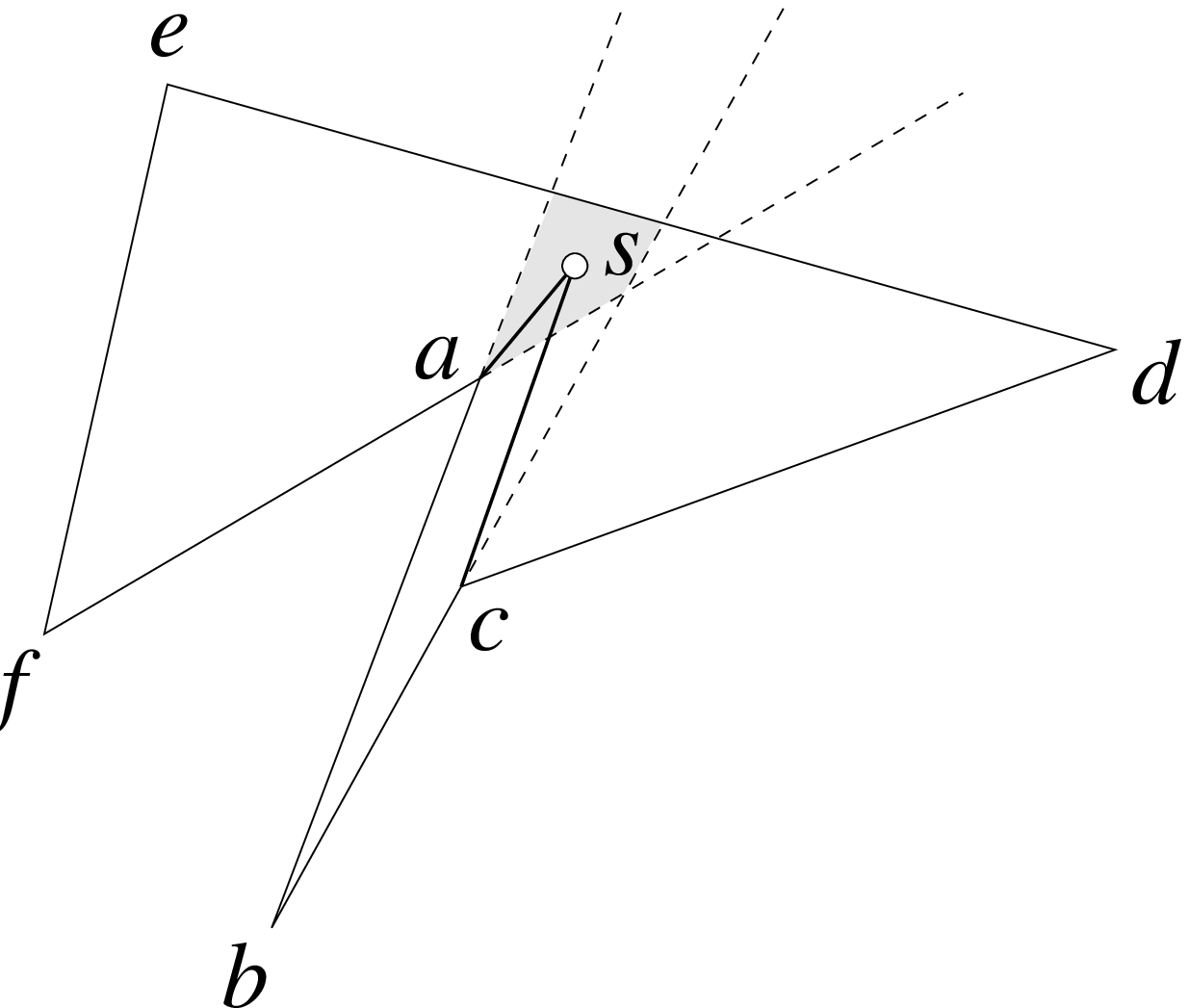}}
      \caption{One Steiner point reduces the problem to the one reflex
        vertex case.} 
      \label{fig:q62_2}

\end{figure}
\item If $d\not\in \Wedge(a)$, then $d$ must lie on one side of
  $\Wedge(a)$%
  \begin{longonly}
    ({i.e.}, $d\in L(b,a)$ or $d\in R(f,a)$),
  \end{longonly}
 and at
  least one of $e$ or $c$, w.l.o.g.~$e$, must lie on the same side (since both $e$ and $c$ are
  convex). 
\begin{enumerate}
  \Case{rccccc{-}1} If $ce \subset P$, by
  Lemma~\ref{lemma:star} one Steiner point is sufficient.
  \Case{rccccc{-}2} If $c$ and $e$ do not see each other, two Steiner
  points are enough.  Placing a Steiner point $s$ in
  $\Wedge(a)$ and connecting it to $a$ and $c$ decomposes the hexagon
  into a quadrangle $\mathit{abcs}$ and a hexagon $\mathit{ascdef}$
  (see Figure~\ref{fig:q61_3}). 
  \begin{longonly}
    The quadrangle is convex: $a$ is
    convex because $s\in \Wedge(a)$.  The vertex $s$ is convex because
    $c\in R(f,a)$ ($c \not\in L(f,a)$, because then $c$ and $e$ would
    see each other) and hence $c\in R(a,s)$.
  \end{longonly}
The hexagon $\mathit{ascdef}$ is as in the previous case $rccccc$-1
($d$
and $f$ necessarily see each other because of our assumption that $d$
and $e$ lie on the same side of $\Wedge(a)$)  
 and hence can be
quadrangulated with one additional Steiner point.
\end{enumerate}
\end{enumerate}

\paragraph{Hexagon with two reflex vertices.}
There are several different cases, depending on the relative positions of the
two reflex vertices in the polygon boundary.
\begin{enumerate}
\Case{rcrccc}
Suppose that the two reflex vertices 
\begin{shortonly}
(w.l.o.g.~$a$ and $c$)   
\end{shortonly}
are separated by a convex vertex
of the polygon.
\begin{longonly}
 Let us assume that $a$ and $c$ are the reflex vertices
of the hexagon $\mathit{abcdef}$. There are two sub-cases.
\end{longonly}
\begin{enumerate}
\Case{rcrccc{-}1}
If both $a$ and $c$ can see $e$, then one Steiner point is enough.
Note that since $e$ is convex $\triangle(ace) \subset \Wedge(e)$.  By
Lemma~\ref{lemma:twowedge} $\Wedge(a) \cap \Wedge(c) \cap
\triangle(ace) \not = \emptyset$.  It follows 
from \eqref{eq:kernwedge},
that the hexagon is starshaped.  We can then apply
Lemma~\ref{lemma:star}.

\Case{rcrccc{-}3}
Otherwise, one of the reflex vertices, w.l.o.g.~$a$, obstructs
the visibility from the other reflex vertex to $e$. We show that 3
Steiner points suffice.  
 By Lemma~\ref{lemma:obstruct}
$\Wedge(a)\cap\Wedge(c) \cap L(a,c) \not=\emptyset$. Place a Steiner
point $s$ in this region and connect it to $a$ and~$c$ (see
Figure~\ref{fig:q62_2}). 
\begin{longonly}
  The quadrangle $\mathit{abcs}$ must be
  convex: $a$ and $c$ are convex because $s$ belongs to their wedges.
  The vertex $s$ is convex because it belongs to $L(a,c)$.
\end{longonly}
The remaining hexagon has only one
reflex vertex $s$, hence can be convex-quadrangulated with at most 2
additional Steiner points.
\end{enumerate}

\Case{rrcccc{-}2}
If the two reflex vertices are consecutive, then two Steiner points
are always sufficient. 
Let $a$ and $b$ be the two reflex vertices.
\begin{longonly}
 Notice that since there
are only two consecutive reflex vertices, $a$ must necessarily 
see $e$, and $b$
must see $d$. Place one Steiner point $s$ in $\Wedge(a)\cap
R(a,e)\cap L(b,d)$. This region is not empty because $de\in L(b,d)$,
and $\Wedge(a)\cap R(a,e)$ contains a subset of the edge $de$ (this can 
be seen by noting that either $d$ or $e$ belong to $\Wedge(a)$, or they 
lie on opposite sides of $\Wedge(a)$).
Connect $s$ to $a$ and $e$ (refer to Figure~\ref{fig:q62_3}).
The quadrangle $\mathit{asef}$ must be convex: $a$ is convex since
$s\in \Wedge(a)$, and $s$ is convex because $s\in R(a,e)$. The
remaining hexagon has two reflex vertices, namely $s$ and $b$,
separated by a convex vertex~$a$. Both $s$ and $b$ can see $d$, since
$s\in L(b,d)$. This is the $\mathit{rcrccc}$-1 case, which requires
one additional Steiner point.
\end{longonly}
\begin{shortonly}
  Placing a Steiner point  $s\in \Wedge(a)\cap
R(a,e)\cap L(b,d)$ and connecting $s$ to $a$ and $e$ (see Figure~\ref{fig:q62_3}) reduces this case 
to the $rcrccc$-1 case.
\end{shortonly}
\begin{figure}[tbhp]                                                        
  \begin{center}

    {\includegraphics[width=.3\columnwidth]{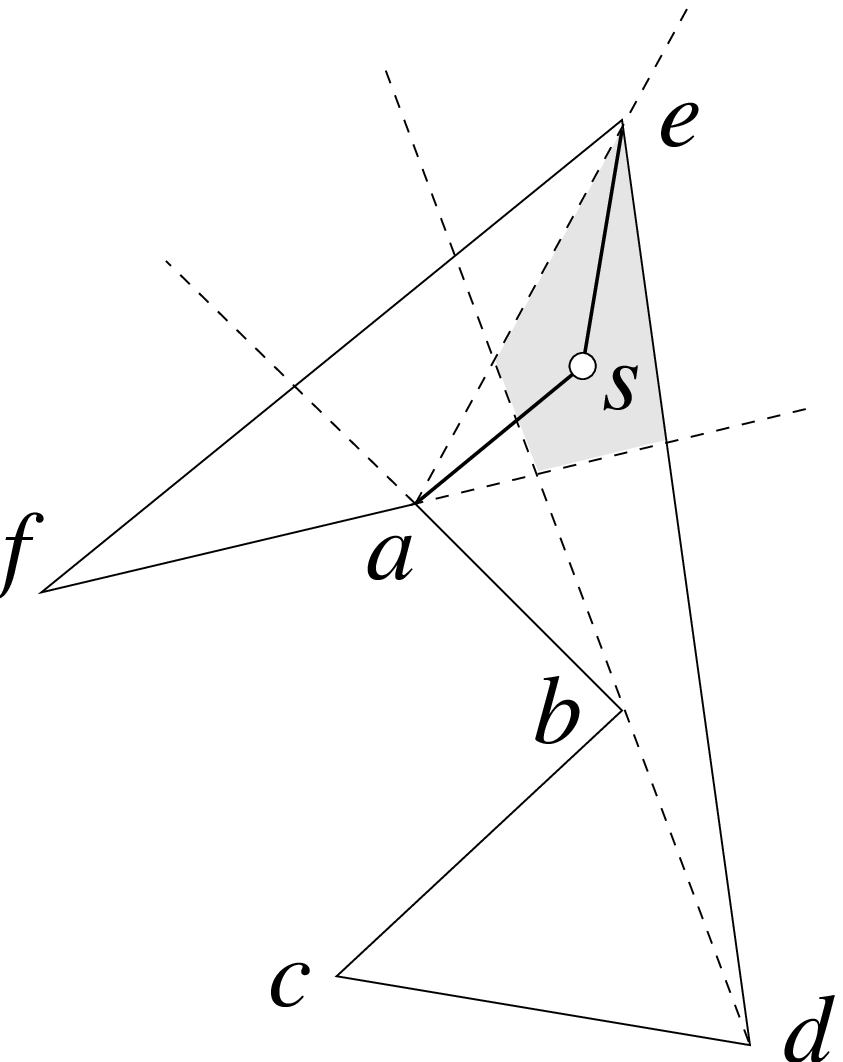}}
    \caption{One Steiner point reduces the problem to the
      $\mathit{rcrccc}$-1 case.} 
    \label{fig:q62_3}
    \end{center}
  \end{figure}

\begin{figure}[tbhp]
\centerline{\includegraphics[width=.4\columnwidth]{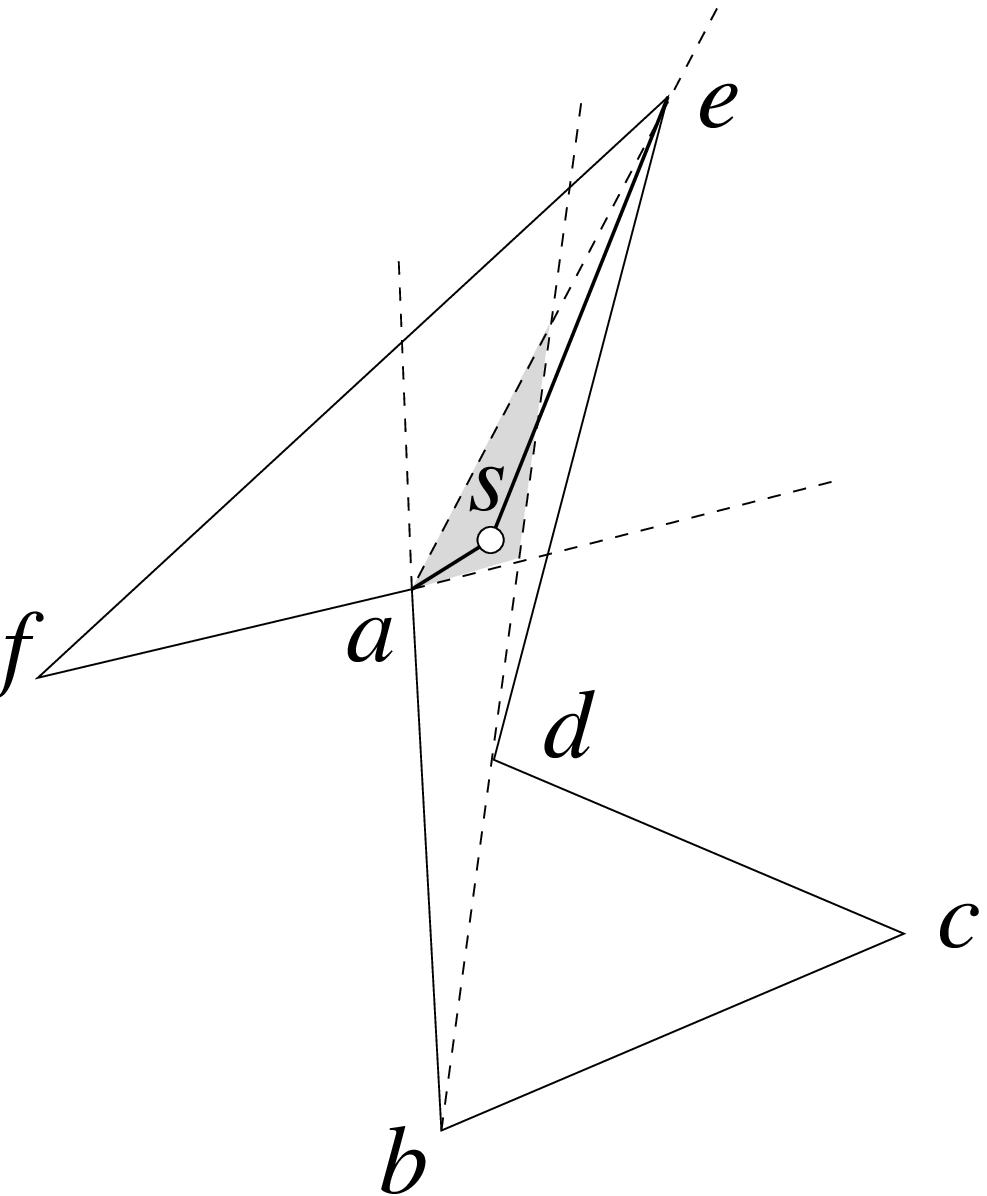}}
\caption{One Steiner point reduces $rccrcc$-2 to  ${rcrccc}$-1
case.}  
\label{fig:q62_4}

\end{figure}

\Case{rccrcc{-}2}
We are left with the case in which there are two convex vertices
between the two reflex vertices, both clockwise and
counterclockwise. 
In this
case, two Steiner points suffice. \label{rccrcc}
Let $a$
and $d$ be the reflex vertices.  
We will use the fact that  
 either the
two diagonals $ae$ and $bd$ are internal to the polygon or $ac$ and
$df$ are. 
  The reason is that if $ac$ is obstructed by $d$, then $d$ belongs to
  $\triangle(abc)$ (recall that $a,b,c,d$ are consecutive) and must
  see $b$, which implies that diagonal $ae$ cannot be obstructed. A
  symmetric argument holds if $df$ is obstructed by~$a$.
Let us assume that $ae$ and $bd$ are internal diagonals (see
Figure~\ref{fig:q62_4}).  Then one Steiner point $s$ can be placed in
$\Wedge(a)\cap R(a,e)\cap L(b,d)$.
\begin{longonly}
This region can be seen to be nonempty as follows:
from the convexity of $f$, $f$ must belong to $L(a,e)$ and hence 
$R(a,e) \cap \Wedge(a) \not= \emptyset$.
In fact $R(a,e) \cap \Wedge(a)$ contains a neighborhood of $a$, which
is in turn contained in $L(b,d)$.  
\end{longonly}
  Connect $s$ to $a$ and $e$. The quadrangle $\mathit{asef}$ is convex.
The remaining polygon is the $\mathit{rcrccc}$-1 type: $s$ and $d$ are
its reflex vertices, and they both see $b$, since $s\in L(b,d)$.
\end{enumerate}

\paragraph{Hexagon with three reflex vertices.}
Again, there are different situations, depending on the relative
positions of the reflex vertices along the polygon boundary.

\begin{enumerate}
\Case{rcrcrc} We start with the case in which the reflex and the convex
  vertices alternate. 

  \begin{enumerate}
  \Case{rcrcrc{-}1} In the special case that $\triangle(ace)$ is 
inside the polygon \emph{and} the polygon is star shaped, 
Lemma ~\ref{lemma:star} implies that one Steiner point suffices. 
    \Case{rcrcrc{-}3} Otherwise, we show that 3 Steiner points suffice. 
    The region $\rho=\Wedge(a) \cap \Wedge(e) \cap R(a,e)$ must be 
    non-empty for the following reason:  
    If $\triangle(ace)$ is inside the polygon, then $\rho$ is
    non-empty as a consequence of Lemma~\ref{lemma:twowedge}. If on the
    other hand one of the edges of $\triangle(ace)$, w.l.o.g.\ $ac$ is
    obstructed, then $\rho$ is non-empty by Lemma~\ref{lemma:obstruct}.
    Place a Steiner point $s$ inside $\rho$.
    Connect $s$ to $a$ and $e$. 
    The quadrangle $efas$ is convex.  
      Vertices $a$ and $e$ are convex by virtue of $s$ being in the
      appropriate wedges.  The vertex $s$ is convex because
      $s\in R(a,e)$.
    The hexagon $sabcde$ is of type $rccrcc$-2 (since $s\in \Wedge(a)
    \cap \Wedge(e)$) hence can be quadrangulated with two additional
    Steiner points.
  \end{enumerate}

\Case{rrcrcc}
We now study the case in which there are exactly two consecutive
reflex vertices. These polygons are always star-shaped,%
\begin{longonly}
for the
following reasons: 
Suppose that $a$, $b$ and $d$ are the reflex vertices (refer to
Figure \ref{fig:q63_3}).
Consider the wedges of $f$ and $c$. The point $e$ must lie on the left
ray of $\Wedge(f)$, and to the right of (or on) the right ray of
$\Wedge(c)$ (since $d$ is reflex). As a consequence, these two
rays must intersect (inside $P$) in a point that we will call $i$.
Since $d$ is reflex, it must lie in the segment ${ci}$, and $e$
cannot lie in the interior of segment~${fi}$. As a consequence, some
portion of the edge ${ef}$ must belong to
$\Wedge(f)\cap \Wedge(c)\cap \Wedge(e)=\interior\kernel(P)$,
(see \eqref{eq:kernconv}).  
\end{longonly}%
\begin{shortonly}
since if $a$, $b$, and $d$ are the reflex vertices, 
$\Wedge(f)\cap \Wedge(c)\cap \Wedge(e)\not=\emptyset$.
\end{shortonly}
 We have two cases depending on
whether $e$ sees at least one of $a$ and~$b$.

\begin{longonly}
  \begin{figure}[tbhp]
    \centerline{\includegraphics[width=0.4\columnwidth]{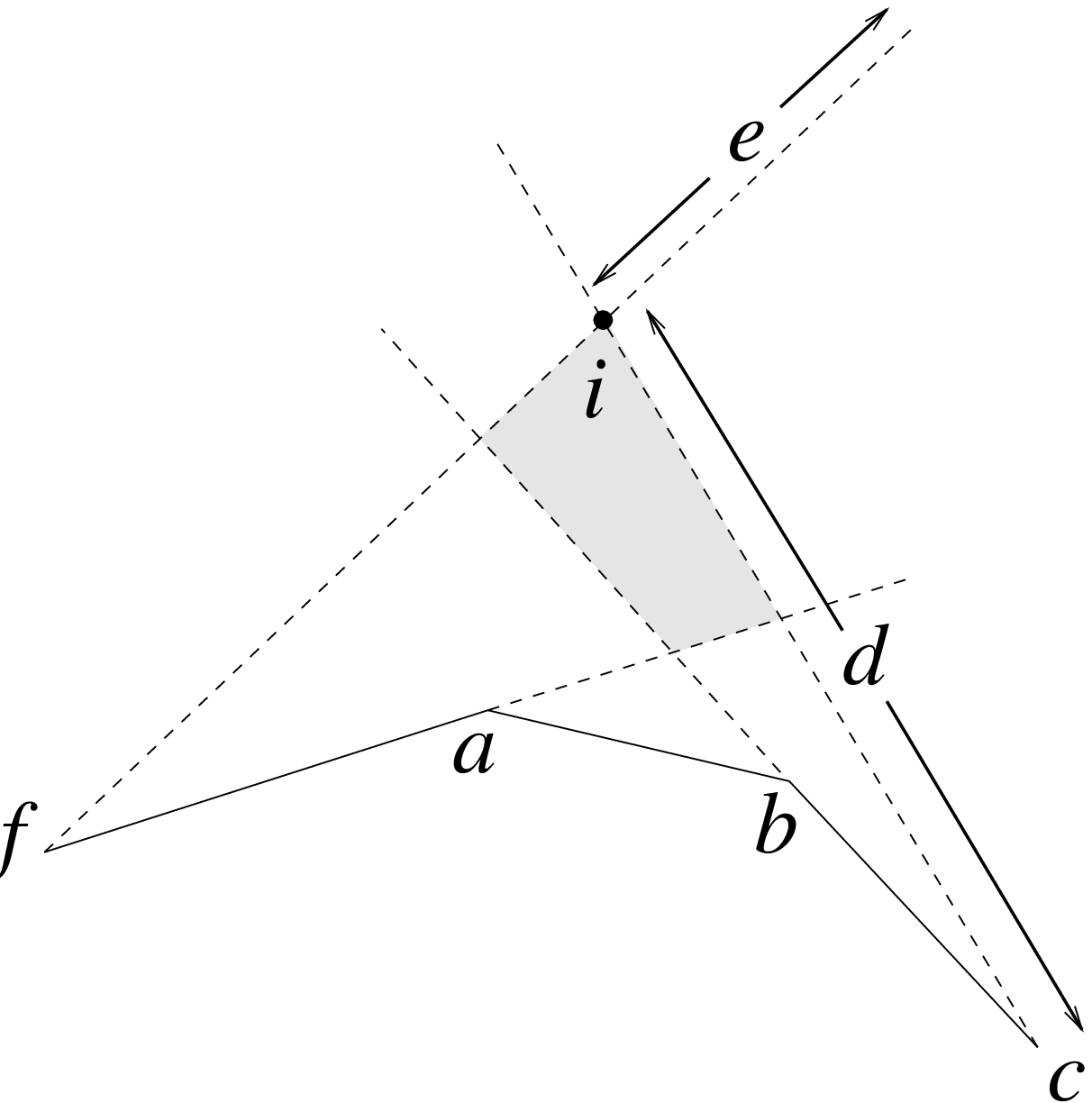}}
\caption{Proving that the $\mathit{rrcrcc}$ polygons are starshaped.}
\label{fig:q63_3}
  \end{figure}
\end{longonly}
\begin{figure}[tbhp]
\centerline{\includegraphics[width=0.5\columnwidth]{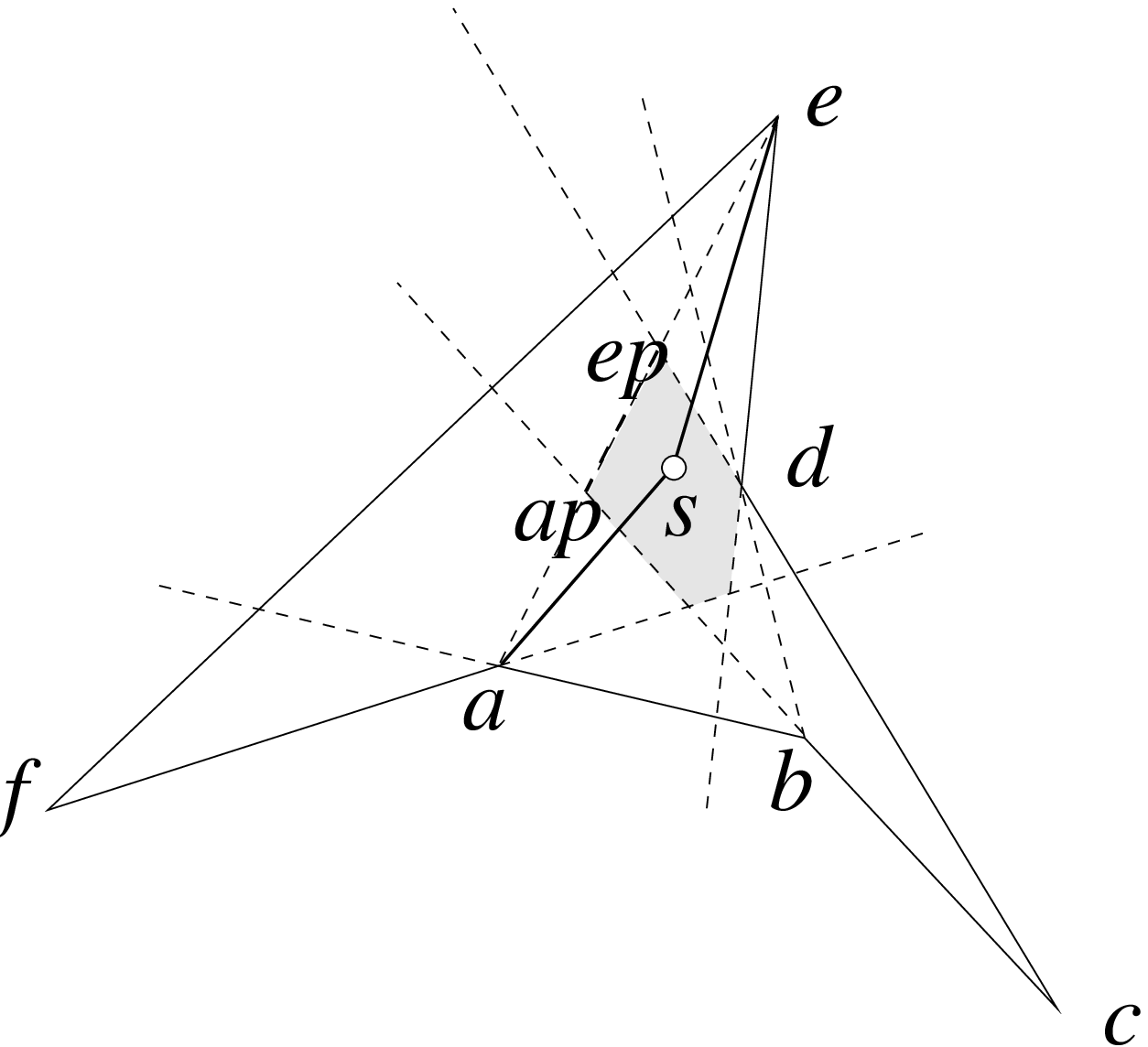}}
\caption{One Steiner point reduces the problem to the
$\mathit{rcrcrc}$-1 case.} 
\label{fig:q63_4}
\end{figure}
\begin{figure}
\centerline{\includegraphics[width=.4\columnwidth ]{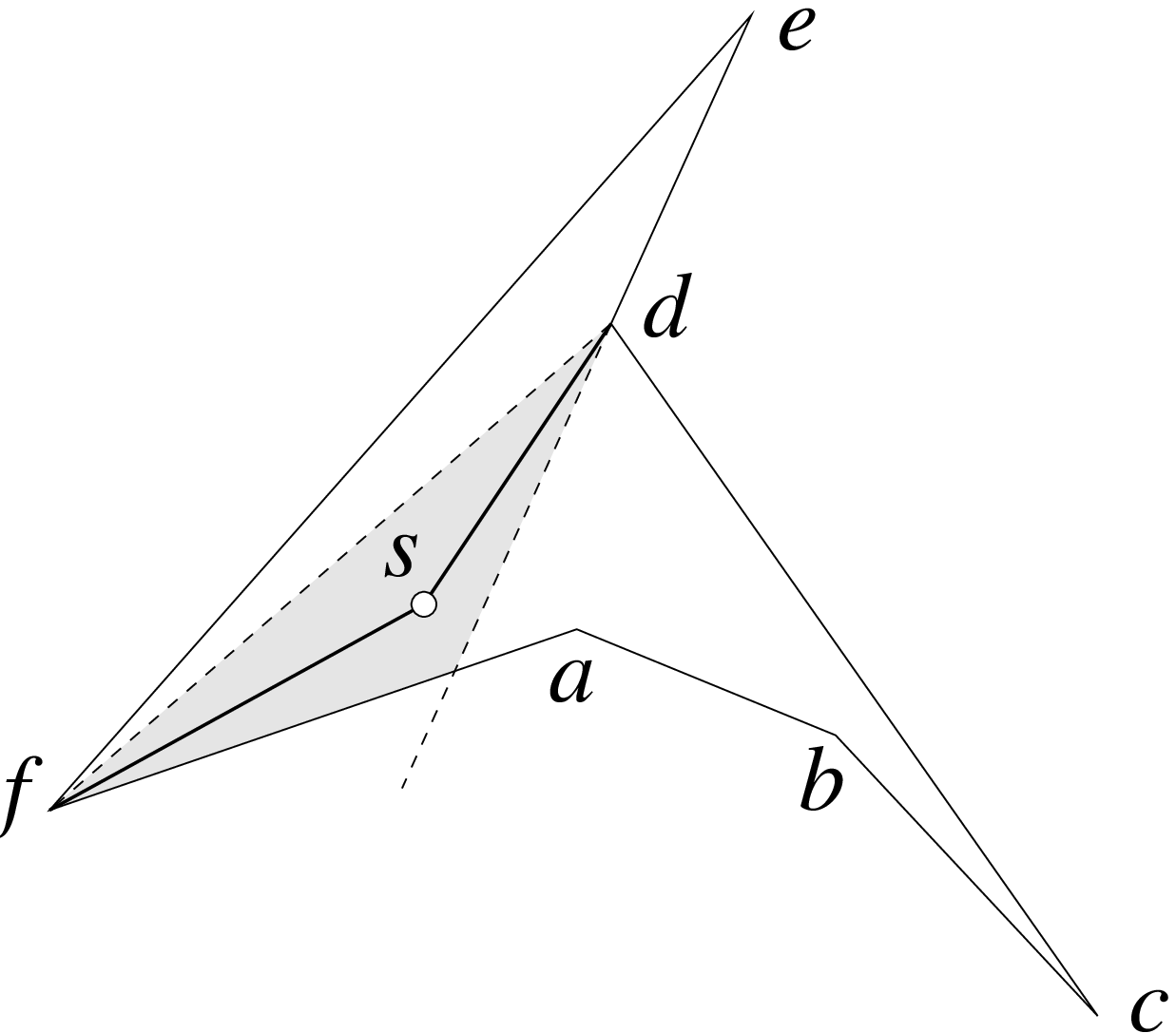}}
\caption{One Steiner point reduces the problem to the
$\mathit{rrcrcc}$-2 case.} 
\label{fig:q63_5}
\end{figure}

\begin{enumerate}
\Case{rrcrcc{-}2}
If $e$ sees at least $a$, two Steiner points suffice.
In particular
the region $\kernel(P) \cap R(a,e) \cap L(b,d)$ (see Figure \ref{fig:q63_4}) cannot be
empty%
\begin{longonly}%
, for the following reason:
The fact that $e$ and $a$ see each other implies that $f\in L(a,e)$
and $b,c,d\in R(a,e)$.
Hence $ae\subset \Wedge(f)\cap \Wedge(e)$. On the other hand,
$\Wedge(c)$ must intersect $ae$, since $e$ lies to its right (because
$d$ is reflex) and similarly $a$ lies to its left. Let $a'e'$ be the
intersection of $\Wedge(c)$ with $ae$ (see Figure~\ref{fig:q63_4}).
Since $a'e'\subseteq\kernel(P)$ and $a'e'\in L(b,d)$ (because $b$
belongs to segment $ca'$ and $d$ belongs to segment $ce'$), it follows
that
$\kernel(P)\cap R(a,e)\cap L(b,d)\not=\emptyset$%
\end{longonly}%
.
Place a Steiner point $s$ in the region, and connect it to $a$ and
$e$.  The quadrangle $\mathit{asef}$ is convex: $a$ is convex
because $s\in \Wedge(a)$, and $s$ is convex because $s\in
R(a,e)$.  The hexagon $\mathit{abcdes}$ is of the
$\mathit{rcrcrc}$-1 type because $s, b, d$ are mutually visible (since
$s\in L(b,d)$).

\Case{rrcrcc{-}3}
If $e$ sees neither $a$ nor $b$, then three Steiner points
suffice. 

\begin{longonly}
In fact, we can reduce the problem to the previous one, after adding
one Steiner point $s$ in the region $\Wedge(e) \cap R(f,d)$
(see Figure
\ref{fig:q63_5}), which must be non-empty.
The point $s$ can then be connected to $f$ and $d$. The quadrangle
$\mathit{sdef}$ is convex: $d$ is convex because $s\in
\Wedge(d)$, and $s$ is convex because $s\in R(f,d)$. The
remaining hexagon is of the kind $\mathit{rrcrcc}$-2, since $d$ can
see both $a$ and $b$, because $s\in \Wedge(e)$.
\end{longonly}
\begin{shortonly}
  Placing a Steiner point $s$ in the region $\Wedge(e) \cap R(f,d)$
  and connecting $s$ to $f$ and $d$ reduces this case to the $rrcrcc$-2 case
(see Figure
\ref{fig:q63_5}).
\end{shortonly}
\end{enumerate}

\begin{longonly}
\begin{figure}[hbtp]               

\begin{center}
{\includegraphics[width=.4\columnwidth]{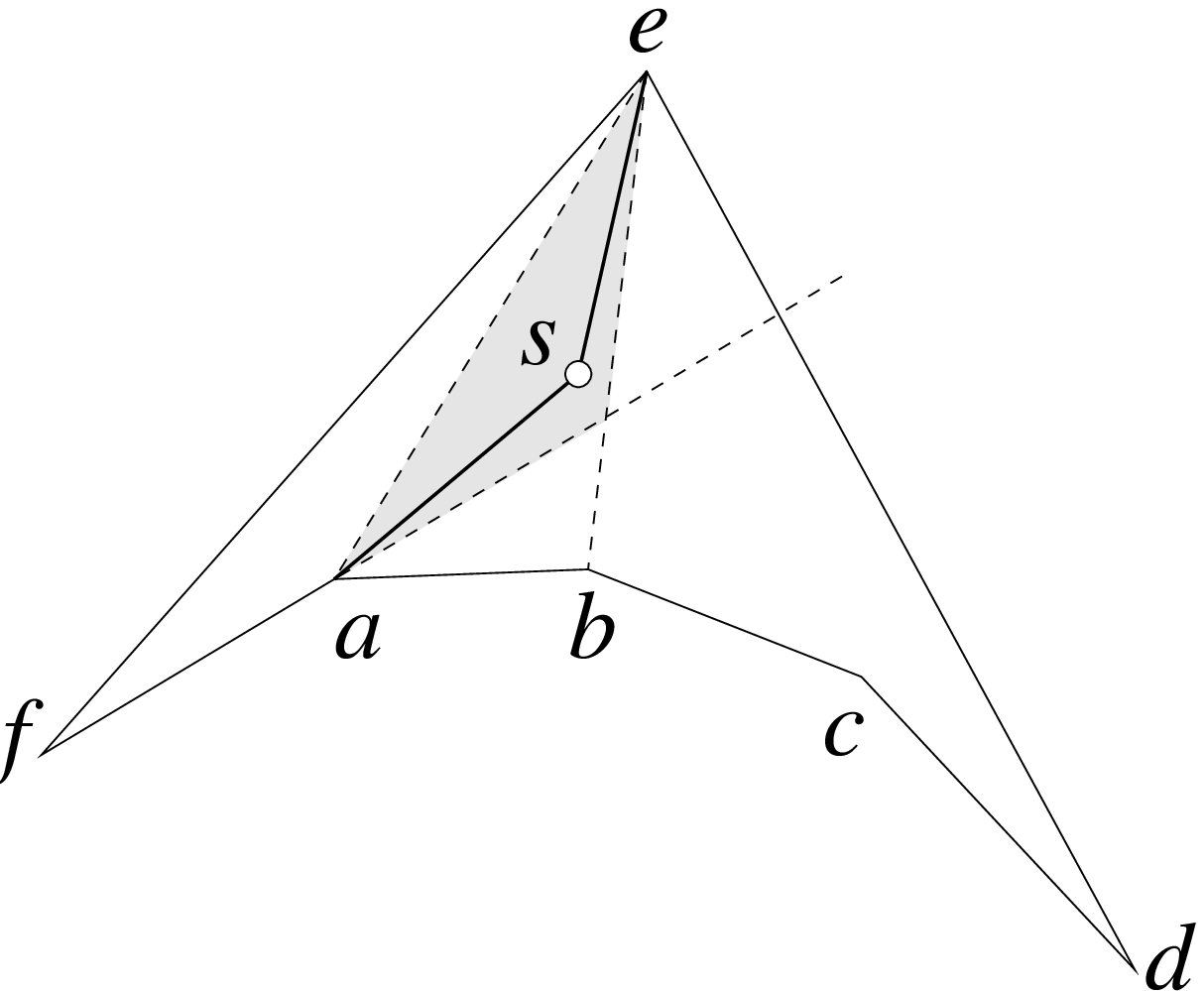}}
\caption{One Steiner point reduces the problem to the
$\mathit{rrcrcc}$-2 case.} 
\label{fig:q63_6}
    \end{center}

\end{figure}
  
\end{longonly}

\Case{rrrccc} We are left with the case in which the three reflex
vertices are consecutive. This case can be solved with three Steiner
points.  In fact, it can be reduced to the $\mathit{rrcrcc}$-2 case
after adding one Steiner point.  Suppose that the three reflex
vertices are $a$, $b$ and $c$.  Place a Steiner point $s$ in the
region $\Wedge(a) \cap R(a,e) \cap
L(b,e)$, which is trivially non-empty. Connecting $s$ with $a$ and
$e$ gives rise to the convex quadrangle $\mathit{asef}$: $a$ is convex
because $s\in \Wedge(a)$, and $s$ is convex because $s\in
R(a,e)$. The remaining hexagon is of the $\mathit{rrcrcc}$-2 type,
since $e$ sees $b$ and $c$, because $s\in L(b,e)$%
\begin{longonly}
(see Figure
\ref{fig:q63_6})  
\end{longonly}%
.
\end{enumerate}

This completes the proof of Theorem~\ref{theorem:hexquad}.
It remains to consider the case when the union of two 
quadrangles is not a hexagon.
\subsubsection{Quadrangle with one interior point.}\label{quadconv}
As stated earlier, when two quadrangles share two edges, their union
is a quadrangle which contains one of the vertices of the original
quadrangles in its interior.  We will show that three Steiner points
suffice to convex-quadrangulate this polygon, thus establishing the 
following theorem:

\begin{theorem}
  Any union of two quadrangles can be convex{-}\relax quadrangulated 
  with at most three Steiner points.
\end{theorem}

\begin{proof}
  We consider here only the case where the union is not a hexagon.
  Let us call the four vertices of the union quadrangle $r$, $a$, $b$
and $c$, where $r$ is the only (possibly) reflex vertex. Let $i$ be
the interior point. Since only $r$ may be reflex, $i$ must see either
$a$ or $c$%
\begin{longonly}
, because $r$ cannot obstruct its view to both%
\end{longonly}%
. Suppose that
$i$ sees $a$, as illustrated in Figure \ref{fig:q5_1}.
\begin{figure}[bthp]                                                        
  \centerline{\includegraphics[width=.4\textwidth]{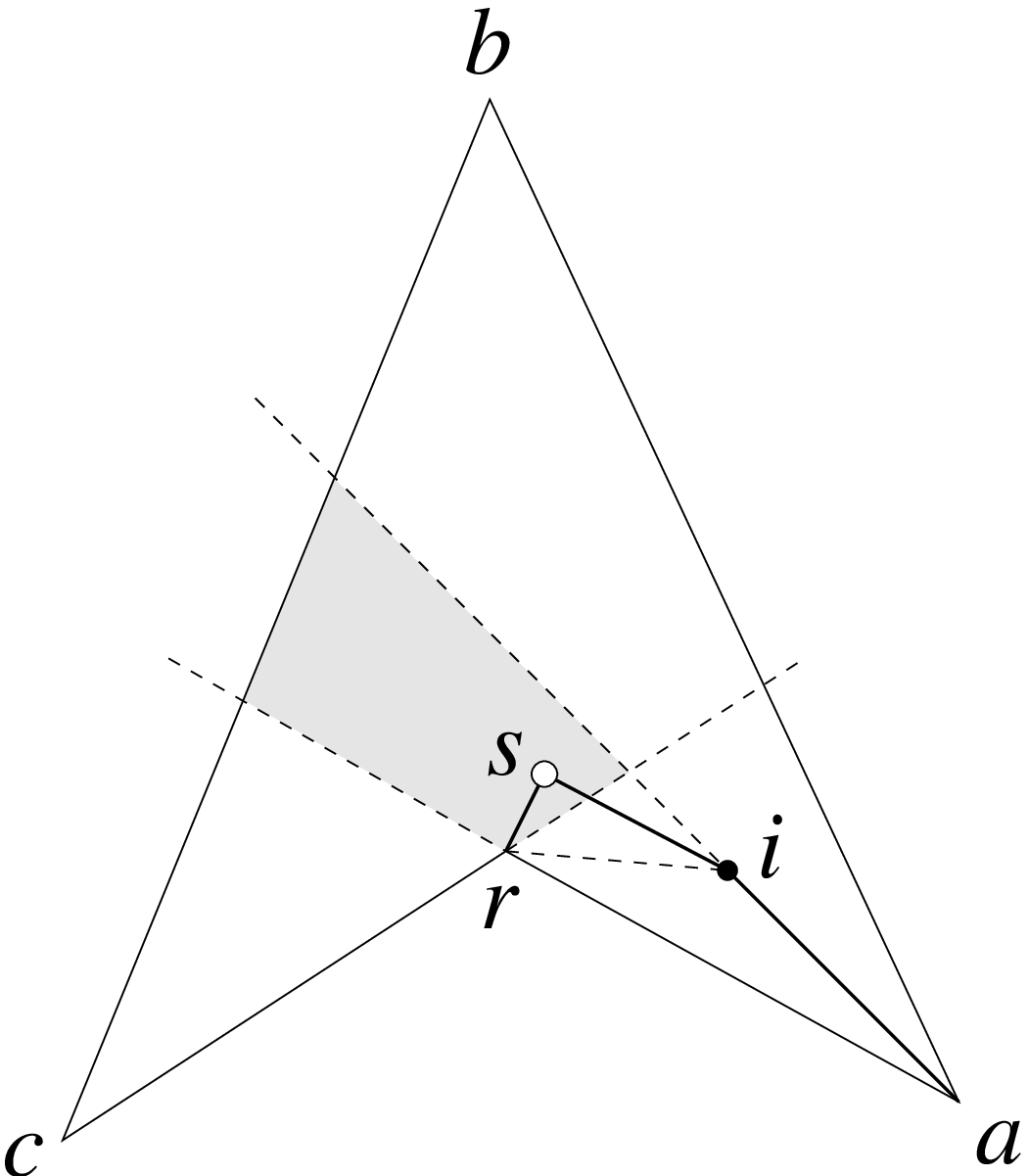}}
  \Caption{One Steiner point reduces the problem to the
    $\mathit{rrcccc}$ case.} 
  {fig:q5_1}
\end{figure}
Since $i\in \Wedge(a)$, 
$\Wedge(r)\cap L(r,i)\cap L(a,i)\not=\emptyset$.  Place one Steiner point
$s$ in the region. Then the quadrangle $\mathit{rais}$ in convex: $r$
is convex because $s\in \Wedge(r)$, $i$ is convex because
$s\in L(a,i)$, and $s$ is convex because $s\in L(r,i)\cap\Wedge(r)$. On the other
hand, the hexagon $\mathit{siabcr}$ is a $\mathit{rrcccc}$ hexagon,
which can be convex-quadrangulated with two Steiner points.
\qed\end{proof}

Each of the cases described in this section runs in constant time, thus:
\begin{theorem}
  A strictly convex quadrilateral mesh of $n$ points using at most
  $3\lfloor{\frac{n}{2}}\rfloor$ Steiner points can be computed in $O(n
  \log n)$ time. 
\end{theorem}

\section{Concluding Remarks\label{sec:concl}}

We have given upper and lower bounds on the number of Steiner points
required to construct a convex quadrangulation for a planar set of
points. Both bounds are constructive, and the upper bound yields 
a straightforward $O(n \log n)$ time algorithm.
The obvious open problem is that of reducing the gap between
the lower and upper bounds. One way to reduce the upper bound may be
by constructing a convex quadrangulation of the point set directly,
rather than by converting a triangulation (by combining triangles
and then quadrangles) as we do now. 
\begin{longonly}
  Also, it would be interesting to
  explore the possibility of improving (raising) the lower bound for a
  non-degenerate point set by combining in some way the two point set 
  configurations given in Section~\ref{sec:lower}.
\end{longonly}

\bibliographystyle{abbrv}
\bibliography{venues,quadran}

\end{document}